\documentclass[conference]{IEEEtran}

\usepackage[ruled,vlined,lined,commentsnumbered]{algorithm2e}
\usepackage{cite}
\usepackage{booktabs}
\usepackage{moreverb}
\usepackage{fontenc}
\usepackage{amsmath}
\usepackage{amssymb,amsfonts}
\usepackage{algorithmic}
\usepackage{fancybox}
\usepackage{color}
\usepackage{colortbl}
\usepackage{array}
\usepackage{multirow}
\usepackage{multicol}
\usepackage{listings}
\usepackage{makecell}
\usepackage{graphicx}
\usepackage{textcomp}
\usepackage{setspace}

\usepackage[ruled,vlined,lined,commentsnumbered]{algorithm2e}
\usepackage{balance}
\usepackage{csvsimple}
\usepackage{longtable}
\usepackage{url}
\usepackage{lscape}
\usepackage{rotating}
\usepackage{tikz}
\usepackage[normalem]{ulem}
\usepackage{enumitem}
 \usepackage{subfigure}

\usepackage[most]{tcolorbox}

\usepackage{threeparttable}

\usepackage{marvosym}
\usepackage{pifont}

\newcolumntype{L}[1]{>{\raggedright\let\newline\\\arraybackslash\hspace{0pt}}m{#1}}
\newcolumntype{C}[1]{>{\centering\let\newline\\\arraybackslash\hspace{0pt}}m{#1}}
\newcolumntype{R}[1]{>{\raggedleft\let\newline\\\arraybackslash\hspace{0pt}}m{#1}}

\definecolor{codegreen}{rgb}{0,0.6,0}
\definecolor{codegray}{rgb}{0.5,0.5,0.5}
\definecolor{codepurple}{rgb}{0.58,0,0.82}
\definecolor{backcolour}{rgb}{0.95,0.95,0.92}

\lstdefinestyle{mystyle}{
    commentstyle=\color{codegreen},
    keywordstyle=\color{magenta},
    numberstyle=\tiny\color{codegray},
    stringstyle=\color{codepurple},
    basicstyle=\footnotesize,
    breakatwhitespace=false,
    breaklines=true,
    captionpos=b,
    keepspaces=true,
    showspaces=false,
    showstringspaces=false,
    showtabs=false,
    tabsize=2
}

\setlist{noitemsep} 

\definecolor{darkpastelred}{rgb}{0.76, 0.23, 0.13}
\definecolor{ao(english)}{rgb}{0.0, 0.5, 0.0}

\definecolor{darkpastelred}{rgb}{0.76, 0.23, 0.13}
\definecolor{ao(english)}{rgb}{0.0, 0.5, 0.0}

\lstdefinelanguage{diff}{
  morecomment=[f][\color{blue}]{@@},     
  morecomment=[f][\color{red}]-,         
  morecomment=[f][\color{codegreen}]+,       
  morecomment=[f][\color{red}]{---}, 
  morecomment=[f][\color{codegreen}]{+++},
}

\definecolor{yellow}{RGB}{255,255,153}
\definecolor{grey}{RGB}{224,224,224}

\newboolean{showcomments}
\setboolean{showcomments}{true}
\ifthenelse{\boolean{showcomments}}
 { \newcommand{\mynote}[2]{
      \fbox{\bfseries\sffamily\scriptsize#1}
        {\small$\blacktriangleright$\textsf{\emph{#2}}$\blacktriangleleft$}}}
        { \newcommand{\mynote}[2]{}}

\definecolor{DarkOrange}{rgb}{0.8,0.3,0.0}
\definecolor{DarkCyan}{rgb}{0.0, 0.55, 0.55}

\newcolumntype{?}{!{\vrule width 1pt}}

\newcommand{\numUnfixedBugs}{246\xspace} 

\newcommand{\toolname}{\texttt{kPAR}\xspace}

\newcommand{\problem}[1]{
\begin{tcolorbox}[tile,size=fbox,boxsep=1mm,boxrule=0pt,top=0pt,bottom=0pt,
borderline west={0.5mm}{0pt}{black!50!white},colback=black!5!white]
\em #1
\end{tcolorbox}
}

\newcommand{\find}[1]{
\begin{tcolorbox}[tile,size=fbox,boxsep=1mm,boxrule=0pt,top=0pt,bottom=0pt,
borderline west={0.6mm}{0pt}{blue!50!white},colback=blue!5!white]
\em #1
\end{tcolorbox}
}

\begin{document}

\title{
You Cannot Fix What You Cannot Find!\\
{\LARGE An Investigation of Fault Localization Bias in}\\
\vspace{-0.3cm}{\LARGE Benchmarking Automated Program Repair Systems}
}

\author{\vspace{-35cm}}
\author{
    \IEEEauthorblockN{
        Kui Liu,
        Anil Koyuncu,
        Tegawend\'e F. Bissyand\'e,
        Dongsun Kim,
        Jacques Klein,
        Yves Le Traon
    }
    \IEEEauthorblockA{Interdisciplinary Centre for Security, Reliability and Trust (SnT), University of Luxembourg, Luxembourg
        \\\{kui.liu, anil.koyuncu, tegawende.bissyande, dongsun.kim,	 jacques.klein, yves.letraon\}@uni.lu
    }
}


\maketitle

\begin{abstract}

Properly benchmarking Automated Program Repair (APR) systems should contribute to the development and adoption of the research outputs by practitioners.
To that end, the research community must ensure that it reaches significant milestones by reliably comparing state-of-the-art tools for a better understanding of their strengths and weaknesses.
In this work, we identify and investigate a practical bias caused by the fault localization (FL) step in a repair pipeline.
We propose to highlight the different fault localization configurations used in the literature,
and their impact on APR systems when applied to the Defects4J benchmark.
Then, we explore the performance variations that can be achieved by ``tweaking'' the FL step. Eventually, we expect to create a new momentum
for (1) full disclosure of APR experimental procedures with respect to FL,
(2) realistic expectations of repairing bugs in Defects4J,
as well as (3) reliable performance comparison among the state-of-the-art APR systems,
and against the baseline performance results of our thoroughly assessed \toolname repair tool.
Our main findings include: (a) only a subset of Defects4J bugs can be currently localized by commonly-used FL techniques;
(b) current practice of comparing state-of-the-art APR systems (i.e., counting the number of fixed bugs) is potentially misleading due to the bias of FL configurations;
and (c) APR authors do not properly qualify their performance achievement with respect to the different tuning parameters implemented in APR systems.
\end{abstract}


\section{Introduction}
\label{sec:intro}

\vspace{-1mm}
Automated program repair (APR) holds the promise of reducing manual debugging effort by automatically generating patches for defects identified in a program.
In production, APR will drastically reduce time-to-fix delays and limit downtime. In a development cycle, APR can help suggest changes to accelerate debugging.
In the literature, there are two distinct repair scenarios: (1) fixing {\em syntactic errors},
i.e., cases where code violates some programming language specifications~\cite{gupta2017deepfix,bhatia2016automated} and (2) fixing {\em semantic bugs}, i.e., cases where implementation of program behaviour deviates from developer's intention~\cite{mechtaev2018semantic,nguyen2013semfix}. The latter requires Fault Localization (FL) through execution of test cases. It is the scope of this paper.

\begin{figure*}[!t]
    \centering
    \includegraphics[width=0.75\linewidth]{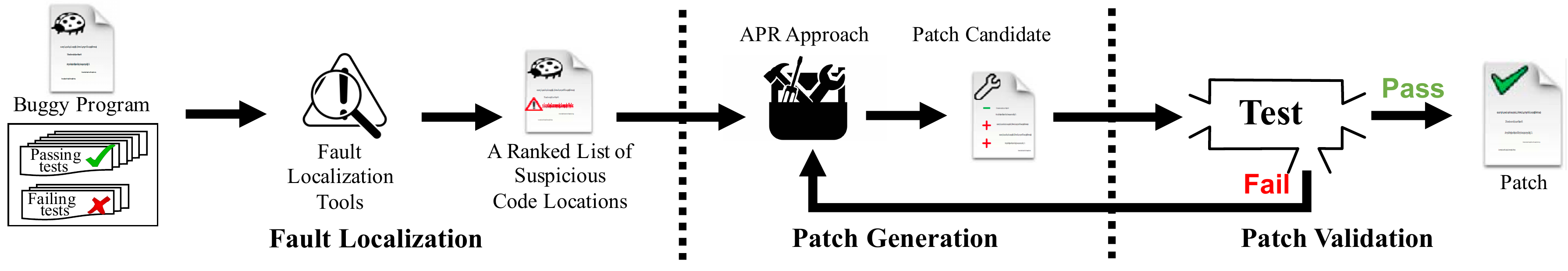}
    \vspace{-2.5mm}
    \caption{Standard steps in a pipeline of Automated Program Repair.}
    \label{fig:APRPro}
\end{figure*}

Once a fault is arosen, most recent APR systems
follow the same basic pipeline as shown in Figure~\ref{fig:APRPro}:
(1) fault localization (FL), (2) patch candidate generation, and (3) patch validation.
The FL step identifies an  entity in a program as the potential fault location.
In patch generation, given a fault location, the APR system
modifies the program, i.e., creates a patch. The last step assesses whether the patch
actually fixes the defect.
If the patch is not regarded as a valid patch, the second and last steps are repeated until a valid patch is generated or the termination condition is satisfied.
To increase the chances of finding a valid patch, the process is iterated over all suspicious code locations ranked by FL tools.

\vspace{-1mm}
In the repair pipeline,
APR systems generally focus on the patch generation step, but tend to
use similar strategies for fault localization and patch validation.
To the best of our knowledge, most of the current state-of-the-art APR approaches~\cite{nguyen2013semfix, westley2009automatically,claire2012genprog,kim2013automatic,coker2013program,ke2015repairing,mechtaev2015directfix,long2015staged,le2016enhancing,xuan2016history,long2016automatic,chen2017contract, le2017s3,long2017automatic,xuan2017nopol,xiong2017precise,liu2019avatar}
leverage test suites to perform fault localization and patch validation.
For fault localization, the systems rely on a testing framework such as GZoltar~\cite{campos2012gzoltar}, and a spectrum-based fault localization formula~\cite{zhang2011non, xuan2014learning, wong2016survey}, such as Ochiai~\cite{abreu2007accuracy}. Eventually, bug fixing performance is measured by counting the number of bugs for which the system can generate a patch that passes all test cases. Such patches are claimed to be valid.

\vspace{-1mm}
Nevertheless, given the growing interest in APR among software engineers, it is important to ensure that the research outputs are relevant and well assessed in terms of reliable performance for practitioners.
In this respect, the APR research community has already started to reflect on the {\em acceptability}~\cite{kim2013automatic,monperrus2014critical} and {\em correctness}~\cite{smith2015cure,xiong2018identifying} of the patches generated by APR tools.
Researchers~\cite{smith2015cure,qi2015analysis,yang2017better,bohme2017bug,le2018overfitting} raised the concern of overfitting patches:
those are generated patches that can pass the validating test cases, but may actually not be the semantically-correct patches for repairing the defect.

\begin{table}[!b]
	\scriptsize
	\centering
	\vspace{1.5mm}
	\setlength\tabcolsep{2pt}
	\caption{Table excerpted from~\cite{jiang2018shaping} with the caption ``{\em Correct patches generated by different techniques}''.}
	\vspace{-1mm}
	\label{tab:simfixComp}
	\resizebox{.9\linewidth}{!}
    {
    	\begin{threeparttable}
			\begin{tabular}{l|c|C{7mm}|C{7mm}|C{7mm}|C{7mm}|C{8mm}|C{7mm}|c|C{7mm}}
			\toprule
			{\bf Proj.} & {\bf SimFix} & {\bf jGP} & {\bf jKali} & {\bf Nopol} & {\bf ACS} & {\bf HDR} & {\bf ssFix} & {\bf ELIXIR} & {\bf JAID}\\
			\hline
			Chart   &  4  & 0 & 0  & 1  & 2  & -(2)   & 3   & 4   & 2(4)  \\
			Closure &  6  & 0 & 0  & 0  & 0  & -(7)   & 2   & 0   & 5(9)  \\
			Math    &  14 & 5 & 1  & 1  & 12 & -(7)   & 10  & 12  & 1/(7) \\
			Lang    &  9  & 0 & 0  & 3  & 3  & -(6)   & 5   & 8   & 1/(5) \\
			Time    &  1  & 0 & 0  & 0  & 1  & -(1)   & 0   & 2   & 0/(0) \\
			\hline
			\hline
			Total   & 34  & 5 & 1  & 5  & 18 & 13(23) & 20  & 26  & 9/(25)\\
			\bottomrule
		\end{tabular}
		\end{threeparttable}
	}
\end{table}

\vspace{-1mm}
Since then, assessment of APR approaches in the literature attempts to provide information on the number of generated patches that are {\em plausible}
(i.e., they make the programs pass all the test cases) and the number of patches that are {\em correct}
(i.e., they are equivalent to the patches that were actually submitted by the program developers).
Table~\ref{tab:simfixComp} provides an example of assessment results excerpted from the
paper describing SimFix~\cite{jiang2018shaping}, one of the most recent state-of-the-art works on APR that was tested on the Defects4J~\cite{just2014defects4j} programs. Based on data reported in this table, researchers explicitly rank the APR systems, and use this ranking as a validation of new achievements in program repair.

\vspace{-0.5mm}
Unfortunately, our own experience in developing and assessing APR tools
has proven that this comparison is non-trivial, and
could further be largely biased due to a non-consideration of important details regarding the FL step.
Indeed, recall that an APR technique cannot attempt to generate the correct patch unless the FL step can successfully identify the target buggy code locations in a program.
Thus, FL accuracy across repair pipelines can impact, either by boosting or degrading, the performance of an APR system.

\vspace{-0.5mm}
For example, SimFix~\cite{jiang2018shaping} and ACS~\cite{xiong2017precise}, although they have been developed by the same research group, are evaluated on different versions of a fault localization technique without discussing the impact of such a change in the experimental configuration. As another example bias, while most APR techniques simply integrate off-the-shelf fault localization tools in the repair pipelines, in some experiments, such as for HDRepair~\cite{xuan2016history}, its authors make the assumption that the buggy method is known. Unfortunately, this assumption gives an important advantage as the list of suspicious code statements is limited and likely to include the buggy statement, thus leading to overestimation of the performance.

\vspace{-0.5mm}
\problem{Similar to the ``overfitting'' study, which helped to improve the assessment criteria of APR tools, our work aims at highlighting the potential biases in comparing different APR approaches without any consideration of implementation variations of the FL step.}

\vspace{-1.5mm}
Overall, our investigation into the relationship between fault localization  performance and APR tool performance seeks to provide answers to the following research questions (RQs):
\begin{enumerate}
\item[{\bf RQ1}] {\bf\em How do APR systems leverage FL techniques?}
We first investigate FL techniques used in APR systems in the literature.
This reveals which FL tool and formula are integrated for each APR system.
We examine implementation details of each APR system, and/or directly ask the authors of the technique
to clarify FL configuration, e.g., which level of detection granularity is considered, and how many suspicious locations are considered.

%

\item[{\bf RQ2}] {\bf\em How many bugs from a common APR benchmark are actually localizable?}
After aggregating APR performance data reported in the literature, we note that \numUnfixedBugs bugs (in benchmark Defects4J) have not yet been fixed by any state-of-the-art APR tool. Given that researchers scarcely discuss the reasons behind repair misses, we assess, with this research question, our intuition that FL is possibly one of the challenging steps in the repair pipeline.
\item[{\bf RQ3}] {\bf\em To what extent APR performance can vary by adapting different FL configurations?} We implement and make publicly available \toolname, a straightforward fix pattern-based APR system, and record its performance under various configurations to serve as a comparable baseline for future research.
\end{enumerate}

Eventually, we make the following contributions:
\begin{itemize}
	\item We expose a hidden bias throughout the comparison of APR tools in the literature, and present more reliable performance comparisons for current state-of-the-art.
	\item We build and make publicly available an easy-to-configure fault localization toolkit that can be adopted in APR pipelines for Java programs.
	\item We provide a refined benchmark for evaluating the performance of APR systems with respect to those bugs that can actually be localized.
	\item We implement and make publicly available a baseline APR system with its different performance metrics for different FL configurations.
\end{itemize}

Our replication package, including \toolname,
is available at:
\begin{center}
\tt\footnotesize
\url{https://github.com/SerVal-DTF/FL-VS-APR}
\end{center}
%



\begin{table*}[!t]
	\scriptsize
	\centering
	\setlength\tabcolsep{2pt}
	\caption{Number of bugs reported having been fixed by different APR tools. {\em APR systems are ordered by year of publication}.}
	\vspace{-1.2mm}
	\label{tab:comparison}
	\resizebox{1\linewidth}{!}
    {
    	\begin{threeparttable}
			\begin{tabular}{l|c|c|c|c|c|c|c|c|c|c|c|c|c|c}
			\toprule
			 {\bf Proj.}  & {\bf jGenProg}~\cite{martinez2016astor}  & {\bf jKali}~\cite{martinez2016astor}  & {\bf jMutRepair}~\cite{martinez2016astor}  & {\bf HDRepair}~\cite{xuan2016history}  & {\bf Nopol}~\cite{xuan2017nopol}  & {\bf ACS}~\cite{xiong2017precise}  & {\bf ELIXIR}~\cite{saha2017elixir}  & {\bf JAID}~\cite{chen2017contract}  & {\bf ssFix}~\cite{xin2017leveraging}  & {\bf CapGen}~\cite{wen2018context}  & {\bf SketchFix}~\cite{hua2018towards}  & {\bf FixMiner}~\cite{koyuncu2018fixminer}  & {\bf LSRepair}~\cite{kui2018live}  & {\bf SimFix}~\cite{jiang2018shaping} \\
			 \hline
Chart  & 0/7  & 0/6  & 1/4  & 0/2  & 1/6  & 2/2  & 4/7  & 2/4  & 3/7  & 4/4  & 6/8  & 5/8  & 3/8  & 4/8 \\
Closure  & 0/0  & 0/0  & 0/0  & 0/7  & 0/0  & 0/0  & 0/0  & 5/11  & 2/11  & 0/0  & 3/5  & 5/5  & 0/0  & 6/8 \\
Lang  & 0/0  & 0/0  & 0/1  & 2/6  & 3/7  & 3/4  & 8/12  & 1/8  & 5/12  & 5/5  & 3/4  & 2/3  & 8/14  & 9/13 \\
Math  & 5/18  & 1/14  & 2/11  & 4/7  & 1/21  & 12/16  & 12/19  & 1/8  & 10/26  & 12/16  & 7/8  & 12/14  & 7/14  & 14/26 \\
Mockito  & 0/0  & 0/0  & 0/0  & 0/0  & 0/0  & 0/0  & 0/0  & 0/0  & 0/0  & 0/0  & 0/0  & 0/0  & 1/1  & 0/0 \\
Time  & 0/2  & 0/2  & 0/1  & 0/1  & 0/1  & 1/1  & 2/3  & 0/0  & 0/4  & 0/0  & 0/1  & 1/1  & 0/0  & 1/1 \\
\hline
\hline														
Total  & 5/27 & 1/22  & 3/17  & 6/23  & 5/35  & 18/23  & 26/41  & 9/31  & 20/60  & 21/25  & 19/26  & 25/31  & 19/37  & 34/56 \\
\hline
P(\%)  & 18.52 & 4.55 & 17.65  & 26.09  & 14.29  & 78.26  & 63.41  & 29.03  & 33.33  & 84.00  & 73.08  & 80.65  & 51.35  & 60.71  \\
			\bottomrule
		\end{tabular}
		{\footnotesize $^\dagger$ In each column, we provide $x/y$ numbers: $x$ is the number of correctly fixed bugs; $y$ is the number of bugs for which a plausible patch is generated by the APR tool (i.e., a patch that makes the program pass all test cases). The same as other similar tables.}
		\end{threeparttable}
	}
\end{table*}

\vspace{-2mm}
\section{Background}
\label{sec:bg}

We recall how fault localization is important in an APR pipeline,
and describe how current APR systems are assessed.

\vspace{-2mm}
\subsection{Fault Localization in Automated Program Repair}
In APR systems, fault localization (FL) is not only the first step but also seriously affects the performance of the systems. 
Given a buggy program (with its passing and failing test cases), an FL tool is leveraged during the FL step to identify the suspicious buggy code locations as described in Figure~\ref{fig:APRPro}.
The granularity of suspicious locations can be a file, method, or line. Ideally, the location should be both precise and accurate.
If the precision is low (e.g., the granularity is broad such as file), the patch generation step needs to explore a large space of candidate patches.
If the accuracy is low (e.g., the FL step provides a wrong fault location), the subsequent step generates patches for the non-faulty program entity.

Spectrum-based fault localization (SBFL, also referred to as coverage-based fault localization)~\cite{zhang2011non, xuan2014learning, wong2016survey} is one of the most popular FL techniques used in APR systems. This technique
applies a ranking metric to detect faulty code locations by leveraging the execution traces of test cases to calculate the likelihood (based on {\em suspiciousness scores}) of program entities to be faulty. 
The ranking metric is applied to calculate suspiciousness scores for program entities (such as program statements as well as code lines~\cite{pearson2017evaluating}).

In the APR literature~\cite{xiong2017precise, jiang2018shaping, martinez2016astor,hua2018towards, wen2018context}, Ochiai~\cite{abreu2007accuracy} is widely used as the ranking metric of SBFL. Many empirical studies~\cite{xie2013theoretical, steimann2013threats, xuan2014learning} have indeed shown that Ochiai is one of the most effective techniques in localizing the root cause of faults in object-oriented programs. Ochiai computes suspiciousness score of a given source code statement $s$ following the formula of Equation~\ref{eq:ochiai}:
\begin{equation}
\label{eq:ochiai}
\vspace{-1.5mm}
\scriptsize
    S_{ochiai}(s) = \frac{failed(s)}{\sqrt{(failed(s) + passed(s)) * (failed(s) + failed(\neg s)}}
\end{equation}
where $failed(s)$  and $passed(s)$ denote respectively the number of failing and passing tests that executed statement $s$, while $failed(\neg s)$ is the number of failing tests that do not execute statement $s$.
In practice, FL tools eventually report a ranked list of statements associated with the suspiciousness scores.

\vspace{-1mm}
\subsection{APR Performance Assessment}
The current practice of APR studies often evaluates the performance of APR systems based on the number of
successfully fixed bugs~\cite{jiang2018shaping,koyuncu2018fixminer}. We can determine whether a generated patch is successful by counting the number of passing test cases. If a patch can pass all the given test cases (both passing and failing cases given for the buggy version), it is regarded as a successful patch.

However, the number of passing test cases may not correctly assess the effectiveness of generated patches.
Even if a generated patch can pass all test cases, it might break a necessary behavior or introduce other faults, which are not covered by the given test suite~\cite{smith2015cure}.
Moreover, a developer may not accept the patch due to  several reasons such as coding convention~\cite{kim2013automatic,monperrus2014critical}.
These patches are often called {\bf plausible patches} since it needs further investigations
to check whether they are {\bf correct patches} acceptable to developers.
In the literature, {\em correctness} is assessed manually by comparing the generated against the developer-provided patch available in the benchmark.

Similarly, selecting a FL technique could be another issue since it can
make the performance assessment biased.
Our investigations will use Table~\ref{tab:comparison} as a starting point to highlight the problem of FL bias.
This table shows the number of fixed bugs out of the bugs in the Defects4J~\cite{just2014defects4j} benchmark, which are reported by the authors of the current state-of-the-art APR tools in the literature.
The results of jGenProg, jKali and Nopol are extracted from the experimental data reported by Martinez et al.~\cite{martinez2017automatic}. The results of other tools are collected from data reported by papers' authors in the literature.

\section{Experimental Setup}
\label{sec:exp}

Our experiments are based on common tool-support and processes used in the literature. We clarify the experiment design in this section as the basis for understanding the implementation and the conclusions that we draw.

\vspace{-1mm}
\subsection{Definition of Fault Locality}
Although state-of-the-art fault localization tools identify suspicious code lines, this information spans across other code entities such as methods and files, which can be  sufficient for APR mutations. Thus, to compute the performance of fault localization techniques on a benchmark, we consider different granularities of fault locality at the {\em file}, {\em method} and {\em line} levels similar to the fault locality defined by Lucia et al.~\cite{thung2012faults}:

\begin{itemize}[leftmargin=*]
    \item {\bf File}: At this level, we consider that the faulty code is accurately localized if an FL tool reports any line from the buggy code file as suspicious.
    \item {\bf Method}: At this level, we consider that the faulty code is accurately localized if any code line in the buggy method is reported by an FL tool as suspicious.
    \item {\bf Line}: At this level, we consider that the faulty code is accurately localized if suspicious code lines reported by an FL tool contain any of the buggy code lines.
\end{itemize}

\vspace{-1mm}
\subsection{Identification of Correct Fault Locality}
\label{sec:correct:locality}
Our objective is to identify which reported suspicious code position is correct, following the above three levels of fault locality granularity.
In practice, FL tools produce a ranked list of suspicious lines while ground truth data include several code lines as buggy lines as well.
At a given granularity level, if the bug is localized (i.e., there is a match between the suspicious code line and the ground truth fault locations),
we record the associated position of the correct fault locality within the ranked list of suspicious code locations.
Since a bug position could span over several lines, methods, and even over several files, the bug is considered to be correctly localized by an FL tool as long as any reported suspicious code line
can match the ground truth bug locations with the corresponding granularity.

Concretely, we first use the following defintion of bug locations.
The locations of a bug in a faulty program are defined as a {\em bug position set}: $BPos$ = \{$bPos_1$, $bPos_2$, \ldots, $bPos_n$\}, (n $>=$ 1), where $bPos_i$ is a tuple of $(fName$, $Methods$, $Lines)$. For each location, $fName$, $Methods$, and $Lines$ are a file name, a set of methods, and a list of line numbers, respectively, of a bug location. $Methods$ could be $\emptyset$ if the bug is not located in any method in a program. This kind of bugs can be related to a Type Declaration~\cite{typeDeclaration}
or Field Declaration~\cite{fieldDeclaration}
 in Java code.
{\em Math-12} in the Defects4J dataset is an example, which is fixed by inserting an interface {\tt Serializable} into the type declaration~\cite{math12}.


We then check whether a ranked list of suspicious lines by an FL tool can identify bug locations based on the following definition.
Let $SuspL=\{suspL_1, suspL_2, \ldots, suspL_m\}$ be a list of suspicious lines that are reported by an FL tool and ordered by suspiciousness scores.
$suspL_i$ is a tuple of $(fName$, $lineNum$, $rIdx)$, where $lineNum$ is the line number of the code in a file (i.e., $fName$) that is suspected to be the bug location,
and $rIdx$ is the index (i.e., rank) of the line within $SuspL$.
If a suspicious line $suspL_i$ ($i \in [1, m]$) matches any bug location ($BPos$) at a given granularity before other suspicious lines,
it is considered that the FL tool successfully identifies a bug location at the given granularity.
Otherwise, if there is no suspicious line matching a bug location at a given granularity,
the fault is considered as non-localizable at this fault locality granularity.




\subsection{Dataset and Automatic Testing Toolset}
Our study requires execution of fault localization on a reliable dataset.
In this work, we select the Defects4J~\cite{just2014defects4j} dataset as it includes test cases
for buggy Java programs with the associated developer fixes. This dataset has furthermore been used
by all recent state-of-the-art APR systems targeting Java programs.
Table~\ref{tab:defects} summarizes statistics on the number of bugs and test cases available
in the version 1.2.0~\cite{d4j_1_2_0}
of Defects4J that we use in this paper.

\vspace{-1mm}
\begin{table}[!h]
	\centering
	\scriptsize
	\caption{Defects4J dataset information.}
	\vspace{-1mm}
	\resizebox{1\linewidth}{!}{
	\begin{threeparttable}
		\begin{tabular}{l|ccccccc}
			\toprule
			Project & Chart  & Closure & Lang & Math & Mockito & Time & Total \\
			\hline
			\# of bugs & 26 & 133 & 65 & 106 & 38 & 27 & 395\\
			\# of test cases &  2,205 &  7,927 &  2,245 &  3,602 & 1,457 &  4,130 & 21,566 \\
			\bottomrule
		\end{tabular}
		{\footnotesize \# of test cases are excerpted from the Defects4J paper~\cite{just2014defects4j} and~\cite{just2018comparing}.}
	\end{threeparttable}
	}
	\label{tab:defects}
\end{table}

Overall, the dataset includes 395 bugs and 22,954 test cases.
To automate the execution of these test cases for each bug,
we rely on the GZoltar
~\cite{gzoltar,campos2012gzoltar} framework for automatic debugging of Java applications.
GZoltar executes the test cases and produces coverage matrices providing information on
which test cases passed, which failed, which statements were executed when running
each test case, etc. Based on this information, FL techniques can be applied
for ranking suspicious code locations which are likely to be the faulty code.
For the purpose of our study, we have implemented on-top of GZoltar 41 common
ranking metrics~\cite{zhang2011non, xuan2014learning} for fault localization.
Given that Gzoltar has been used by several APR tools in the literature,
we expect that our easy-to-configure fault localization toolkit will serve
the research community to parameterize fault localization in an APR pipeline.

Our experiments further considered two different versions of GZoltar. 
The first one is the GZoltar version 0.1.1, which is already used in state-of-the-art APR systems, such as Astor~\cite{martinez2016astor},
FixMiner~\cite{koyuncu2018fixminer}, ACS~\cite{xiong2017precise}, ssFix~\cite{xin2017leveraging}
and CapGen~\cite{wen2018context} among others.
On the other hand, the GZoltar version
1.6.0 is used in SimFix~\cite{jiang2018shaping} since it was recently shown to be effective~\cite{pearson2017evaluating}.


\subsection{Implementation of a Baseline APR System}
Ideally, we should consider exploring an existing APR system for
drawing our reference performance. Unfortunately, we face several challenges: (1) only a few research groups openly release the code or even implementation details of their APR systems; (2) repair steps are often tightly coupled together in  implementation, which requires substantial engineering effort for experimental adaptation; (3) proposed approaches generally mix several contributions which are hard to isolate.

 We, therefore, propose to implement and share a baseline repair system based on a state-of-the-art publication on Java program repair. We select PAR~\cite{kim2013automatic} for its simplicity and the straightforward replication that can be carried out on the basis of  details from the relevant research report.  We build \toolname, which leverages patterns that have been learned from the commonalities among 60,000 human-written patches. Six common patterns
from the initial version of PAR has been implemented in \toolname.
We further record the performance of \toolname in repair scenarios involving four different configurations of the fault localization step.

\vspace{-2mm}
\section{Study Results}
\label{sec:eval}
We now provide key findings for the related questions that are investigated in this work.

\vspace{-2mm}
\subsection{Integration of FL Tools in APR Pipelines}

To characterize how FL tools are integrated into APR pipelines, we carefully assess evaluation reports in the literature
and investigate the source code (when it is available) of 14 state-of-the-art
APR systems which have been evaluated on the Defects4J benchmark. Table~\ref{tab:aprFLTools} enumerates
the studied tools along with the information collected.
We focus on the testing framework that is used and its version,
the FL ranking metric that is considered to compute the suspiciousness scores,
the granularity of fault locality that authors focused on, and the extra information that authors use to supplement FL.

\begin{table*}[!t]
	\scriptsize
	\centering
	\setlength\tabcolsep{2pt}
	\caption{Fault Localization (FL) techniques integrated into state-of-the-art APR tools.}
	\vspace{-1mm}
	\label{tab:aprFLTools}
	\resizebox{1\linewidth}{!}
    {
    	\begin{threeparttable}
			\begin{tabular}{l|c|c|c|c|c|c|c|c|c|c|c|c|c|c}
			\toprule
			  & {\bf jGP} & {\bf jKali} & {\bf jMutRepair}	& {\bf HDRepair} & {\bf Nopol} & {\bf ACS}  & {\bf ELIXIR} & {\bf JAID}	& {\bf ssFix} & {\bf CapGen} & {\bf SketchFix} & {\bf FixMiner} & {\bf LSRepair}	& {\bf SimFix}\\
			\hline
			\makecell[l]{FL testing framework}   & GZoltar & GZoltar	& GZoltar  & ?  &  GZoltar & GZoltar   & ?   & ?  & GZoltar   & GZoltar & ? & 	GZoltar & GZoltar & GZoltar\\
			\hline
			\makecell[l]{Framework version}& 0.1.1 & 0.1.1	 & 0.1.1 & ?	  & 0.0.10 & 0.1.1  & ? & ? & 0.1.1 & 0.1.1 & ? &	0.1.1 & 0.1.1 	& 1.6.0 \\
			\hline
			\makecell[l]{FL ranking metric}  & Ochiai & Ochiai & Ochiai  & ?   & Ochiai  & Ochiai  &  Ochiai  & ? &  ?  & Ochiai	& Ochiai &  Ochiai	 & Ochiai &  Ochiai \\
			\hline
			\makecell[l]{Granularity of fault locality} & line & line & line & line & line & line & line & line & line & line & line &line & method& line \\
			\hline
			\makecell[l]{Supplementary\\information}  & $\emptyset$ & $\emptyset$ & $\emptyset$ & \makecell[c]{Faulty method\\ is known} & $\emptyset$ & \makecell[c]{Predicate\\switching~\cite{zhang2006locating}} & ? & ? & \makecell[c]{Statements in \\crashed stack trace} & ? & ? & $\emptyset$ & $\emptyset$  & \makecell[c]{Test Case\\Purification~\cite{xuan2014test}}  \\
			\bottomrule
		\end{tabular}
		{\scriptsize $^\ast$ The unspecified/unconfirmed information of an APR tools is marked with `?'. If an APR tool does not use any supplementary information for FL, the corresponding table cell is marked with `$\emptyset$'.}
		\end{threeparttable}
	}
\end{table*}

Among the 14 APR tools that are investigated, 10 leverage GZoltar as the automated testing toolset in the repair pipeline.
Except for SimFix, which uses a recent version of the framework,
all others use earlier versions (8 tools use version 0.1.1, while Nopol uses an even
older version, i.e., 0.0.1). Thus, unless otherwise stated,
the experiments in this work are performed on the widely used version 0.1.1 of GZoltar.

Eleven out of the 14 APR tools are explicitly known to rely on Ochiai for computing the suspiciousness scores in the fault localization process.
This popularity of Ochiai is backed up by empirical evidence on its effectiveness to help localize faults in object-oriented programs as highlighted by several
fault localization studies~\cite{steimann2013threats,xie2013theoretical,xuan2014learning, papadakis2015metallaxis}.
A recent work by Pearson et al.~\cite{pearson2017evaluating} has even shown that
Ochiai outperforms current state-of-the-art ranking metrics, or at least offers similar performance measures.
In the latter part of this study, we replicate their work to ensure that our implementation of the ranking techniques is reliable.
It should also be noted that although ELIXIR and SketchFix do not
report the test framework that they use, they explicitly mention using Ochiai for fault localization.

With respect to the granularity of fault locality, only LSRepair~\cite{kui2018live} focuses on the method-level granularity to detect and fix bugs.
Other APR systems require information on bugs at the line level to proceed with patch generation.
Considering methods as the granularity of fault locations implies that such faults that are
located outside methods (e.g., type declaration faults~\cite{kui2018closer})
will not be addressed. However, this granularity may offer a time advantage:
when several statements in a single method are reported as suspicious locations,
LSRepair, unlike other APR systems, is not required to iteratively try
each location for generating patch candidates.
Finally, it should be noted that FL tools do not offer the same accuracy
in identifying faulty locations at different granularity levels
(cf. Section~\ref{subsec:localizability}), making method level granularity appealing
for limiting unnecessary trials on fault positive locations.

It is further noteworthy that four APR systems leverage supplementary information
to assist the fault localization step and improve accuracy. The impact of this
assistance is unfortunately never discussed when comparing performance among
state-of-the-art repair approaches.
Typically:
\begin{itemize}[leftmargin=*]
	\item HDRepair~\cite{xuan2016history} assumes that the faulty methods are known:
	the fault localization step therefore focuses on ranking the lines inside the method,
	thus leaving out noisy statements that other APR tools are considering.
	This artificially reduces the probability to produce overfitting patches for
	HDRepair, and even increases the chance to generate a correct patch before any execution timeout.
	\item ssFix~\cite{xin2017leveraging} prioritizes statements from the stack trace
	of crashed programs that are executed before those statements that are ranked by the FL tool.
	\item ACS~\cite{xiong2017precise} uses predicate switching~\cite{zhang2006locating}
	and refines the suspicious code locations list since the repair is focused on faulty conditional statements.
	\item SimFix~\cite{xuan2014test} applies a test case purification approach to improve
	the accuracy of FL step before patch generation.
\end{itemize}
Although these extra steps, which are taken to supplement FL step, could be justified intuitively, the community needs to clearly investigate their impact,
in order to enable fair comparisons among the repair techniques themselves.
Indeed, given that APR systems are currently compared with respect to the number
of bugs that are correctly fixed, it is important that the research community
reflects on what are the key contributions for explaining APR performance:
for example, by counting numbers of correct patches, several programs may not be repairable
by a given APR system simply because the fault is not accurately localized by the implemented FL step.

\find{{\bf RQ1}$\blacktriangleright$State-of-the-art APR systems in the literature add some adaptations to the usual FL process to improve its accuracy.
Unfortunately, researchers have eluded so far the contribution of this improvement
in the overall repair performance, leading to biased comparisons.}

\subsection{Localizability of Defects4J Bugs}
\label{subsec:localizability}
\vspace{-1.5mm}
In a recent work, Koyuncu et al.~\cite{koyuncu2018fixminer} have reported
that 136 bugs in total from the Defects4J dataset have already been associated
to a plausible patch that was generated by at least one APR system from the literature.
Patches for 83 bugs have even been validated as correct patches by researchers.
Considering this data that we complement with the performance realized by another
recent APR tool, namely LSRepair, we conclude that $\sim$62\% (246/395) of Defects4J's
bugs have never seen a plausible patch automatically generated by the state-of-the-art in APR.
Although a recent empirical study~\cite{motwani2018automated} has suggested that current
APR systems cannot repair hard and important bugs, our intuition is that there might
be a more practical issue related to the localizability of Defects4J defects:
\begin{center}
\vspace{-1.5mm}
{\em How many faults in the Defects4J benchmark can actually be localized by
current automated fault localization tools?}
\vspace{-1.5mm}
\end{center}



We consider the most common scenario of fault localization scenario from the APR literature: GZoltar is used for automated test execution, and Ochiai for computing suspiciousness scores. Test execution is performed  with the test cases provided in the Defects4J benchmark. Table~\ref{tab:detectedBugs} provides quantitative details on the localizability of bugs at different levels of fault locality granularity (i.e., file, method and line). Experiments are performed with two distinct versions of GZoltar.

\begin{table}[!b]
	\scriptsize
    \centering
    \caption{Number of Bugs localized$^\ast$ with Ochiai/GZoltar.}
    \label{tab:detectedBugs}
    \begin{threeparttable}
    \begin{tabular}{l|C{9mm}|C{6mm}|C{6mm}|C{6mm}|C{6mm}|C{6mm}|C{6mm}}
        \toprule
        \multirow{2}{*}{\bf Project} & \multirow{2}{*}{\bf \# Bugs} & \multicolumn{2}{c|}{\bf File} & \multicolumn{2}{c|}{\bf Method} & \multicolumn{2}{c}{\bf Line} \\\cline{3-8}
        & & {\bf GZ$_1$} & {\bf GZ$_2$}& {\bf GZ$_1$} & {\bf GZ$_2$}& {\bf GZ$_1$} & {\bf GZ$_2$} \\\cline{1-8}
        Chart   & 26  & 25 & 25  & 22 & 24 & \cellcolor[gray]{0.8}22 & 24 \\
        Closure & 133 & 113 & 128 & 78 & 96 & \cellcolor[gray]{0.8}78 & 95 \\
        Lang    & 65  & 54 & 64  & 32 & 59 & \cellcolor[gray]{0.8}29 & 57 \\
        Math    & 106 & 101 & 105 & 92 & 100 & \cellcolor[gray]{0.8}91 & 100 \\
        Mockito & 38  & 25 & 26  & 22 & 24 & \cellcolor[gray]{0.8}21 & 23 \\
        Time    & 27  & 26 & 26  & 22 & 22 & \cellcolor[gray]{0.8}22 & 22 \\
        \hline
        Total   & 395 & 344& 374 & 268 & 325 & \cellcolor[gray]{0.8}263 & 321 \\
        \bottomrule
    \end{tabular}
    {\footnotesize $^\ast$A bug is counted as localized as long any of the faulty locations appear in the ranked list of suspicious locations reported by the FL tool. GZ$_1$ and GZ$_2$ indicate GZoltar 0.1.1 and 1.6.0, respectively. The same abbreviations are used for GZoltar versions in the following tables.
		The column GZ$_1$ of ``Line'' is highlighted since it is the most common configuration in APR systems.}
    \end{threeparttable}
\end{table}

In this experiment, we consider a bug to be localized as long as the faulty code is listed among the suspicious statements reported by this fault localization tools.
Considering the most common configuration in the literature (GZoltar version 0.1.1 and ``Line''
granularity level), up to 132 (= 395 - 263) bugs in Defects4J are not localized.
The number of bugs that are not localized decreases to 74 (= 395 - 321) when
the coverage matrices are produced with GZoltar version 1.6.0.
This result suggests that with GZoltar version 1.6.0, APR systems have
an opportunity attempt the fix of 58 more bugs.

\find{{\bf RQ2}$\blacktriangleright$One third of bugs in the Defects4J dataset cannot
be localized by the commonly used automated fault localization tool.
Nevertheless, the recent version of GZoltar provides coverage information
that helps localize more than 50 bugs, which may have never been considered
in validation trials of early APR systems.}


\vspace{-1.5mm}
Besides Ochiai, we have attempted to localize bugs in the Defects4J benchmark
by using six other ranking metrics to compute suspiciousness scores.
Table~\ref{tab:flCannot} presents the number of bugs localized by the different ranking metrics.
We consider the cases where the actual fault location is  reported at
the Top-1 position of the suspicious code locations, and among the Top-10 positions.
Results for Top-50, Top-100, Top-200 and all localized are also made available in the replication package.
The results show that fault localization performance is consistent among the different ranking metrics.

\begin{table}[!h]
    \centering
    \vspace{-1.5mm}
    \scriptsize
    \setlength\tabcolsep{2pt}
    \caption{Number of Bugs localized at Top-1 and Top-10.}
    \vspace{-0.5mm}
    \label{tab:flCannot}
    \begin{threeparttable}
    \begin{tabular}{L{16mm}|C{10mm}|C{10mm}|C{10mm}||C{10mm}|C{10mm}|C{10mm}}
        \toprule
       \multirow{2}{*}{\makecell[l]{{\bf Ranking}\\{\bf Metric}}}  & \multicolumn{3}{c||}{\bf GZ$^1$} & \multicolumn{3}{c}{\bf GZ$^2$} \\\cline{2-7}
	& {\bf File} & {\bf Method} & {\bf Line} & {\bf File}  & {\bf Method} & {\bf Line} \\
	\hline
 \multicolumn{7}{l}{\em Top-1 Position} \\ \cline{1-7}
Tarantula & 171 & 101 & 45 & 169 & 106 & 35 \\
\rowcolor[gray]{0.8}Ochiai & 173 & 102 & 45 & 172 & 111 & 38 \\
DStar2 & 173 & 102 & 45 & 175 & 114 & 40 \\
Barinel & 171 & 101 & 45 & 169 & 107 & 36 \\
Opt2 & 175 & 97 & 39 & 179 & 115 & 39 \\
Muse & 170 & 98 & 40 & 178 & 118 & 41 \\
Jaccard & 173 & 102 & 45 & 171 & 112 & 39 \\
\hline
 \multicolumn{7}{l}{\em	 Top-10 Position} \\ \cline{1-7}
Tarantula & 240 & 180 & 135 & 242 & 189 & 144 \\
\rowcolor[gray]{0.8}Ochiai & 244 & 184 & 140 & 242 & 191 & 145 \\
DStar2 & 245 & 184 & 139 & 242 & 190 & 142 \\
Barinel & 240 & 180 & 135 & 242 & 190 & 145 \\
Opt2 & 237 & 168 & 128 & 239 & 184 & 135 \\
Muse & 234 & 169 & 129 & 239 & 186 & 140 \\
Jaccard & 245 & 184 & 139 & 241 & 188 & 142 \\
        \bottomrule
    \end{tabular}
    \end{threeparttable}
\end{table}

Only 45 bugs can be accurately localized with Ochiai at the first suspicious line location. 140 and 214 bugs can be localized at Top-10 and Top-100 positions. Actually, many APR systems only focus on generating patches iteratively based on a part of the list of suspicious code locations.
For example, for SketchFix~\cite{hua2018towards}, authors explicitly declare to consider only the top-50 most suspicious statements in the ranked list, while in ELIXIR~\cite{saha2017elixir}, up to the top-200  suspicious locations are considered.

\subsection{Impact of Effective Ranking in Fault Localization}
Automated fault localization produces a ranked list of suspicious code locations that APR tools must iteratively consider for patch generation.
To assess to what extent effective ranking (i.e., placing the actually faulty code locations at the top of the list), we propose to investigate the
correlation between the rank of bug localization in the suspicious lists and the ability of state-of-the-art systems to be able to repair it.

Table~\ref{tab:results.apr} summarizes the list of all bugs, from the Defects4J benchmark, for which a plausible patch has been generated by one of the 14 state-of-the-art APR systems considered in this study.
For each bug, we indicate the rank of the bug location within the ranked list of suspicious locations provided by the fault localization for different localization granularities.
Experiments are done using the Ochiai ranking metric, but with two versions of GZoltar for computing the test coverage matrices. The raw data, including for other ranking metrics, are available in our replication package.

\begin{table*}
	\caption{Localization positions (i.e., rank within the suspicious list) for Defects4J bugs which have been fixed (correctly or plausibly) by corresponding APR systems.}
	\label{tab:results.apr}
	\scriptsize
	\setlength\tabcolsep{1.5pt}
	\begin{tabular}{cc}
		\begin{minipage}{.50\textwidth}
			\resizebox{1\linewidth}{!}{
				\begin{tabular}{|l|c|c|c|c|c|c|c|c|c|c|c|c|c|c?c|c|c?c|c|c?c|c|c?c|c|c?c|c|c?c|c|c?c|c|c|}
\hline
& \multicolumn{14}{c?}{APR tools}	& \multicolumn{3}{c?}{{GZ$^1$ \& Ochiai}} & \multicolumn{3}{c?}{{GZ$^2$ \& Ochiai}} \\
\hline
\rotatebox[origin=l]{0}{Bug ID}   &   \rotatebox[origin=l]{90}{jGenProg}   &   \rotatebox[origin=l]{90}{jKali}   &   \rotatebox[origin=l]{90}{jMutRepair}   &   \rotatebox[origin=l]{90}{HDRepair}   &  \rotatebox[origin=l]{90}{Nopol}   &   \rotatebox[origin=l]{90}{ACS}   &   \rotatebox[origin=l]{90}{ELIXIR}   &   \rotatebox[origin=l]{90}{JAID}   &   \rotatebox[origin=l]{90}{ssFix}   &   \rotatebox[origin=l]{90}{CapGen}   &   \rotatebox[origin=l]{90}{SketchFix}   &   \rotatebox[origin=l]{90}{FixMiner}   &   \rotatebox[origin=l]{90}{LSRepair}   &   \rotatebox[origin=l]{90}{SimFix}   &   \rotatebox[origin=l]{90}{File}                     &   \rotatebox[origin=l]{90}{Method}                     &   \rotatebox[origin=l]{90}{Line}                      &   \rotatebox[origin=l]{90}{File}                     &   \rotatebox[origin=l]{90}{Method}                    &   \rotatebox[origin=l]{90}{Line}                \\
 \hline 
     Chart-1   &          \ding{109}   &       \ding{109}   &            \ding{108}   &          \ding{109}   &  &  &        \ding{108}   &    (\ding{108})   &       \ding{108}   &        \ding{108}   &           \ding{108}   &          \ding{108}   &          \ding{108}   &        \ding{108}   & 1 & 1 & 28 & 24 & 24 &         28 \\
     Chart-3   &          \ding{109}   &  &  &  &       \ding{109}   &  &        \ding{109}   &  &  &  &  &  &  &        \ding{108}   & 1 & 1 & 7 & 1 & 1 &          4 \\
     Chart-4   &  &  &  &  &  &  &  &  &  &  &  &          \ding{108}   &          \ding{108}   &  & 1 & 1 & 49 & 2 & 2 &        173 \\
     Chart-5   &          \ding{109}   &       \ding{109}   &  &  &       \ding{108}   &  &  &  &       \ding{109}   &  &  &  &  &  & 1 & 1 & 7 & 12 & 66 &         72 \\
     Chart-6   &  &  &  &  &  &  &  &  &  &  &  &  &          \ding{109}   &  & 2 & 49 & 49 & 24 & 222 &        224 \\
     Chart-7   &          \ding{109}   &  &            \ding{109}   &  &  &  &  &  &       \ding{109}   &  &  &  &  &        \ding{108}   & 1 & 2 & 28 & 1 & 2 &         75 \\
     Chart-8   &  &  &  &          \ding{109}   &  &  &        \ding{108}   &  &  &        \ding{108}   &           \ding{108}   &  &  &  &         {\bf \color{red}0}  &         {\bf \color{red}0}  &         {\bf \color{red}0}  & 1 & 1 &          1 \\
     Chart-9   &  &  &  &  &  &  &        \ding{108}   &    (\ding{108})   &       \ding{109}   &  &           \ding{108}   &  &  &  & 1 & 1 & 3 & 1 & 2 &         14 \\
    Chart-10   &  &  &  &  &  &  &  &  &  &  &  &  &          \ding{109}   &  & 1 & 1 & 1 & 1 & 3 &          3 \\
    Chart-11   &  &  &  &  &  &  &        \ding{108}   &  &  &        \ding{108}   &           \ding{108}   &          \ding{108}   &          \ding{108}   &  & 1 & 1 & 15 & 1 & 24 &         28 \\
    Chart-12   &  &  &  &  &  &  &  &  &  &  &  &          \ding{109}   &  &        \ding{109}   & 1 &         {\bf \color{red}0}  &         {\bf \color{red}0}  & 801 & 1092 &       1093 \\
    Chart-13   &          \ding{109}   &       \ding{109}   &  &  &       \ding{109}   &  &        \ding{109}   &  &       \ding{109}   &  &           \ding{109}   &          \ding{109}   &  &  & 1 & 1 & 17 & 5 & 35 &         51 \\
    Chart-14   &  &  &  &  &  &     \ding{108}   &  &  &  &  &  &  &  &        \ding{109}   & 1 & 1 & 1 & 38 & 996 &        998 \\
    Chart-15   &          \ding{109}   &       \ding{109}   &  &  &  &  &  &  &  &  &  &  &  &  & 1 & 26 & 26 & 2143 & 8444 &       8445 \\
    Chart-17   &  &  &  &  &  &  &        \ding{109}   &  &  &  &  &  &          \ding{109}   &  & 1 & 2 & 2 & 11 & 12 &         12 \\
    Chart-18   &  &  &  &  &  &  &  &  &  &  &  &  &          \ding{109}   &        \ding{109}   & 1 & 1 & 6 & 1 & 1 &          3 \\
    Chart-19   &  &  &  &  &  &     \ding{108}   &  &  &  &  &  &  &  &  & 1 & 1 & 5 & 37 & 833 &        833 \\
    Chart-20   &  &  &  &  &  &  &  &  &       \ding{108}   &  &           \ding{108}   &  &  &        \ding{108}   & 4 &         {\bf \color{red}0}  &         {\bf \color{red}0}  & 62 & 62 &         62 \\
    Chart-21   &  &  &  &  &       \ding{109}   &  &  &  &  &  &  &  &  &  & 2 & 2 & 2 & 1 & 34 &         39 \\
    Chart-22   &  &  &  &  &  &  &  &  &  &  &  &  &  &        \ding{109}   & 1 & 1 & 1 & 1 & 53 &         58 \\
    Chart-24   &  &  &  &  &  &  &  &      \ding{108}   &       \ding{108}   &        \ding{108}   &           \ding{108}   &          \ding{108}   &  &  & 1 & 1 & 2 & 1 & 1 &          3 \\
    Chart-25   &          \ding{109}   &       \ding{109}   &            \ding{109}   &  &       \ding{109}   &  &  &  &  &  &  &          \ding{109}   &  &  & 1 & 30 & 47 & 1325 & 3668 &       3913 \\
    Chart-26   &  &       \ding{109}   &            \ding{109}   &  &       \ding{109}   &  &  &      \ding{108}   &  &  &           \ding{109}   &          \ding{108}   &          \ding{109}   &  & 132 & 132 & 132 & 241 & 14795 &         15053\\
    \hline 
   Closure-5   &  &  &  &  &  &  &  &      \ding{109}   &  &  &  &  &  &  & 8 &         {\bf \color{red}0}  &         {\bf \color{red}0}  & 561 &         {\bf \color{red}0}  &         {\bf \color{red}0}\\
   Closure-7   &  &  &  &  &  &  &  &  &       \ding{109}   &  &  &  &  &  & 7 &         {\bf \color{red}0}  &         {\bf \color{red}0}  & 28 &         {\bf \color{red}0}  &         {\bf \color{red}0}\\
  Closure-10   &  &  &  &          \ding{109}   &  &  &  &  &  &  &  &          \ding{108}   &  &  & 3 & 56 & 120 & 3 & 67 &        141 \\
  Closure-12   &  &  &  &  &  &  &  &  &       \ding{109}   &  &  &  &  &  & 154 & 368 & 368 & 393 & 1085 &       1085 \\
  Closure-14   &  &  &  &          \ding{109}   &  &  &  &  &       \ding{108}   &  &           \ding{108}   &  &  &        \ding{108}   & 1 & 2 & 3 & 2 & 3 &          3 \\
  Closure-18   &  &  &  &  &  &  &  &      \ding{108}   &  &  &  &  &  &  & 90 & 1495 & 1527 & 93 & 2320 &       2377 \\
  Closure-31   &  &  &  &  &  &  &  &    (\ding{108})   &  &  &  &  &  &  & 215 & 1026 & 1043 & 214 & 1756 &       1802 \\
  Closure-33   &  &  &  &  &  &  &  &      \ding{108}   &  &  &  &  &  &  & 2 & 2 & 289 & 2 & 2 &        318 \\
  Closure-38   &  &  &  &  &  &  &  &  &  &  &  &          \ding{108}   &  &  & 1 & 1 & 34 & 1 & 1 &         49 \\
  Closure-40   &  &  &  &  &  &  &  &      \ding{108}   &  &  &  &  &  &  & 9 &         {\bf \color{red}0}  &         {\bf \color{red}0}  & 104 &         {\bf \color{red}0}  &         {\bf \color{red}0}\\
  Closure-42   &  &  &  &  &  &  &  &  &       \ding{109}   &  &  &  &  &  & 4 &         {\bf \color{red}0}  &         {\bf \color{red}0}  & 15 &         {\bf \color{red}0}  &         {\bf \color{red}0}\\
  Closure-51   &  &  &  &          \ding{109}   &  &  &  &  &  &  &  &  &  &  & 25 & 25 & 33 & 41 & 41 &         50 \\
  Closure-57   &  &  &  &  &  &  &  &  &  &  &  &  &  &        \ding{108}   & 1 & 2 & 3 & 1 & 2 &          7 \\
  Closure-62   &  &  &  &          \ding{109}   &  &  &  &    (\ding{108})   &  &  &           \ding{108}   &          \ding{108}   &  &        \ding{108}   & 1 & 1 & 1 & 1 & 1 &          4 \\
  Closure-63   &  &  &  &  &  &  &  &    (\ding{108})   &  &  &  &          \ding{108}   &  &        \ding{108}   & 1 & 1 & 1 & 1 & 1 &          4 \\
  Closure-68   &  &  &  &  &  &  &  &  &       \ding{109}   &  &  &  &  &  & 2 & 2 & 2 & 2 & 2 &          4 \\
  Closure-70   &  &  &  &          \ding{109}   &  &  &  &      \ding{108}   &  &  &           \ding{109}   &  &  &  & 143 &         {\bf \color{red}0}  &         {\bf \color{red}0}  & 264 &         {\bf \color{red}0}  &         {\bf \color{red}0}\\
  Closure-73   &  &  &  &          \ding{109}   &  &  &  &      \ding{108}   &  &  &           \ding{109}   &          \ding{108}   &  &        \ding{108}   & 1 & 7 & 10 & 1 & 1 &         16 \\
  Closure-79   &  &  &  &  &  &  &  &  &  &  &  &  &  &        \ding{109}   &         {\bf \color{red}0}  &         {\bf \color{red}0}  &         {\bf \color{red}0}  & 1 & 37 &         37 \\
 Closure-106   &  &  &  &  &  &  &  &  &  &  &  &  &  &        \ding{109}   &         {\bf \color{red}0}  &         {\bf \color{red}0}  &         {\bf \color{red}0}  & 3 & 4 &          4 \\
 Closure-109   &  &  &  &  &  &  &  &  &       \ding{109}   &  &  &  &  &  & 1 & 9 & 9 & 1 & 4 &          4 \\
 Closure-111   &  &  &  &  &  &  &  &  &       \ding{109}   &  &  &  &  &  & 1 &         {\bf \color{red}0}  &         {\bf \color{red}0}  & 7 &         {\bf \color{red}0}  &         {\bf \color{red}0}\\
 Closure-115   &  &  &  &  &  &  &  &  &       \ding{108}   &  &  &  &  &        \ding{108}   & 1 & 1 & 1 & 8 & 8 &          8 \\
 Closure-122   &  &  &  &  &  &  &  &  &       \ding{109}   &  &  &  &  &  & 1 & 1 & 2 & 1 & 2 &          2 \\
 Closure-125   &  &  &  &  &  &  &  &      \ding{109}   &       \ding{109}   &  &  &  &  &  & 4 & 142 & 145 & 5 & 166 &        170 \\
 Closure-126   &  &  &  &          \ding{109}   &  &  &  &    (\ding{108})   &       \ding{109}   &  &           \ding{108}   &  &  &  & 1 & 1 & 1 & 6 & 6 &          6 \\
 \hline 
      Lang-2   &  &  &  &  &  &  &  &  &       \ding{109}   &  &  &  &          \ding{109}   &  & 127 & 127 & 128 & 1 & 1 &         17 \\
      Lang-6   &  &  &  &          \ding{108}   &  &  &        \ding{108}   &  &       \ding{108}   &        \ding{108}   &           \ding{108}   &  &  &  &         {\bf \color{red}0}  &         {\bf \color{red}0}  &         {\bf \color{red}0}  & 54 & 73 &         74 \\
      Lang-7   &  &  &  &  &  &     \ding{108}   &  &  &  &  &  &  &  &  & 373 &         {\bf \color{red}0}  &         {\bf \color{red}0}  & 1 & 1 &         25 \\
     Lang-10   &  &  &  &          \ding{109}   &  &  &  &  &  &  &  &  &  &        \ding{109}   & 114 &         {\bf \color{red}0}  &         {\bf \color{red}0}  & 1 & 64 &         64 \\
     Lang-16   &  &  &  &  &  &  &  &  &  &  &  &  &  &        \ding{108}   & 322 &         {\bf \color{red}0}  &         {\bf \color{red}0}  & 1 & 1 &         27 \\
     Lang-21   &  &  &  &  &  &  &  &  &       \ding{108}   &  &  &  &          \ding{108}   &  & 1 &         {\bf \color{red}0}  &         {\bf \color{red}0}  & 1 & 1 &          2 \\
     Lang-24   &  &  &  &  &  &     \ding{108}   &        \ding{108}   &      \ding{109}   &  &  &  &  &          \ding{108}   &  & 259 & 618 &         {\bf \color{red}0}  & 1 & 14 &         64 \\
     Lang-26   &  &  &  &  &  &  &        \ding{108}   &  &  &        \ding{108}   &  &  &  &  & 21 &         {\bf \color{red}0}  &         {\bf \color{red}0}  & 1 & 112 &        112 \\
     Lang-27   &  &  &            \ding{109}   &  &  &  &  &  &       \ding{109}   &  &  &  &  &        \ding{108}   & 262 & 269 &         {\bf \color{red}0}  & 1 & 1 &         56 \\
     Lang-29   &  &  &  &  &  &  &  &  &  &  &  &  &          \ding{108}   &  & 59 & 59 &         {\bf \color{red}0}  & 1 & 1 &         {\bf \color{red}0}\\
     Lang-33   &  &  &  &  &  &  &        \ding{108}   &      \ding{108}   &       \ding{108}   &  &  &  &  &        \ding{108}   & 14 &         {\bf \color{red}0}  &         {\bf \color{red}0}  & 1 & 1 &          7 \\
     Lang-35   &  &  &  &  &  &     \ding{108}   &  &  &  &  &  &  &  &  & 53 &         {\bf \color{red}0}  &         {\bf \color{red}0}  & 1 & 1 &          2 \\
     Lang-38   &  &  &  &  &  &  &        \ding{108}   &    (\ding{108})   &  &  &  &  &  &  & 104 &         {\bf \color{red}0}  &         {\bf \color{red}0}  & 1 & 3 &          3 \\
     Lang-39   &  &  &  &  &       \ding{109}   &     \ding{109}   &        \ding{109}   &      \ding{109}   &       \ding{109}   &  &  &  &  &        \ding{108}   & 203 &         {\bf \color{red}0}  &         {\bf \color{red}0}  & 1 & 2 &         27 \\
     Lang-40   &  &  &  &  &  &  &  &  &  &  &  &  &          \ding{109}   &  & 1 & 1 & 1 & 1 & 1 &          2 \\
     Lang-41   &  &  &  &  &  &  &  &  &  &  &  &  &          \ding{109}   &        \ding{108}   & 1 & 5 & 7 & 1 & 5 &          6 \\
     Lang-43   &  &  &  &          \ding{109}   &  &  &        \ding{108}   &  &       \ding{108}   &        \ding{108}   &  &          \ding{109}   &  &        \ding{108}   & 1 & 1 & 1 & 1 & 26 &         29 \\
     Lang-44   &  &  &  &  &       \ding{108}   &  &        \ding{109}   &  &       \ding{109}   &  &  &  &  &        \ding{109}   & 1 & 5 & 20 & 1 & 1 &          3 \\
     Lang-45   &  &  &  &  &  &  &  &    (\ding{108})   &  &  &  &  &  &        \ding{109}   & 1 & 1 & 16 & 1 & 1 &          5 \\
     Lang-46   &  &  &  &  &       \ding{109}   &  &  &  &  &  &  &  &          \ding{108}   &  & 1 & 1 & 3 & 1 & 1 &          1 \\
     Lang-48   &  &  &  &  &  &  &  &  &  &  &  &  &          \ding{108}   &  & 1 & 1 & 2 & 1 & 1 &          2 \\
     Lang-50   &  &  &  &  &  &  &  &  &  &  &  &  &  &        \ding{108}   & 1 & 9 & 15 & 1 & 8 &          8 \\
     Lang-51   &  &  &  &          \ding{108}   &       \ding{109}   &  &        \ding{109}   &    (\ding{108})   &       \ding{109}   &  &           \ding{109}   &  &          \ding{108}   &  & 1 & 1 &         {\bf \color{red}0}  & 1 & 1 &         {\bf \color{red}0}\\
     Lang-52   &  &  &  &  &  &  &  &  &  &  &  &  &          \ding{108}   &  & 1 & 1 & 13 & 1 & 3 &         25 \\ 
     Lang-53   &  &  &  &  &       \ding{109}   &  &  &  &  &  &  &  &  &  & 1 & 1 & 32 & 1 & 1 &         16 \\
     Lang-54   &  &  &  &  &  &  &  &  &  &  &  &  &          \ding{108}   &  & 1 & 1 & 2 & 1 & 1 &          4 \\
Lang-55   &  &  &  &  &       \ding{108}   &  &  &    (\ding{108})   &  &  &           \ding{108}   &  &          \ding{109}   &  & 1 & 7 & 9 & 1 & 6 &          7 \\
     Lang-57   &  &  &  &          \ding{109}   &  &  &        \ding{108}   &  &  &        \ding{108}   &  &          \ding{108}   &  &  & 1 & 1 & 1 & 1 & 1 &          1 \\
     \hline
\end{tabular}

			}
		\end{minipage}
	&
		\begin{minipage}{.5\textwidth}
			\resizebox{1\linewidth}{!}{
				\begin{tabular}{|l|c|c|c|c|c|c|c|c|c|c|c|c|c|c?c|c|c?c|c|c?c|c|c?c|c|c?c|c|c?c|c|c?c|c|c|}
\hline
& \multicolumn{14}{c?}{APR tools}   & \multicolumn{3}{c?}{{GZ$^1$ \& Ochiai}} & \multicolumn{3}{c?}{{GZ$^2$ \& Ochiai}} \\
\hline
\rotatebox[origin=l]{0}{Bug ID}   &   \rotatebox[origin=l]{90}{jGenProg}   &   \rotatebox[origin=l]{90}{jKali}   &   \rotatebox[origin=l]{90}{jMutRepair}   &   \rotatebox[origin=l]{90}{HDRepair}   &   \rotatebox[origin=l]{90}{Nopol}    &   \rotatebox[origin=l]{90}{ACS}   &   \rotatebox[origin=l]{90}{ELIXIR}   &   \rotatebox[origin=l]{90}{JAID}   &   \rotatebox[origin=l]{90}{ssFix}   &   \rotatebox[origin=l]{90}{CapGen}   &   \rotatebox[origin=l]{90}{SketchFix}   &   \rotatebox[origin=l]{90}{FixMiner}   &   \rotatebox[origin=l]{90}{LSRepair}   &   \rotatebox[origin=l]{90}{SimFix}   &   \rotatebox[origin=l]{90}{File}                     &   \rotatebox[origin=l]{90}{Method}                     &   \rotatebox[origin=l]{90}{Line}                      &   \rotatebox[origin=l]{90}{File}                     &   \rotatebox[origin=l]{90}{Method}                    &   \rotatebox[origin=l]{90}{Line}                \\
\hline  
     Lang-58   &  &  &  &  &       \ding{108}   &  &        \ding{109}   &  &       \ding{109}   &  &  &  &  &        \ding{108}   & 1 & 2 & 8 & 1 & 1 &         20 \\
     Lang-59   &  &  &  &          \ding{109}   &  &  &        \ding{108}   &  &       \ding{108}   &        \ding{108}   &           \ding{108}   &          \ding{108}   &  &  & 1 & 1 & 1 & 1 & 1 &          6 \\
     Lang-60   &  &  &  &  &  &  &  &  &  &  &  &  &          \ding{109}   &        \ding{108}   & 1 & 1 & 3 & 1 & 1 &          2 \\
     Lang-61   &  &  &  &  &  &  &  &      \ding{109}   &  &  &  &  &  &  & 1 & 14 & 19 & 1 & 14 &         21 \\
     Lang-62   &  &  &  &  &  &  &  &  &  &  &  &  &          \ding{109}   &  & 1 & 1 & 7 & 1 & 1 &          7 \\
     Lang-63   &  &  &  &  &  &  &  &  &       \ding{109}   &  &  &  &  &        \ding{109}   & 1 & 1 & 1 & 1 & 1 &          1 \\
    \hline  
      Math-1   &  &  &  &  &  &  &  &  &  &  &  &  &  &        \ding{109}   & 20 &         {\bf \color{red}0}  &         {\bf \color{red}0}  & 12 & 14 &         14 \\
      Math-2   &          \ding{109}   &       \ding{109}   &            \ding{109}   &  &  &  &        \ding{109}   &  &       \ding{109}   &  &  &  &  &  & 1 & 11 & 11 & 44 & 44 &         44 \\
      Math-3   &  &  &  &  &  &     \ding{108}   &  &  &       \ding{109}   &  &  &  &  &  & 1 & 1 & 16 & 1 & 1 &          3 \\
      Math-4   &  &  &  &  &  &     \ding{108}   &  &  &  &  &  &  &  &  & 1 & 1 & 1 & 1 & 3 &          6 \\
      Math-5   &          \ding{108}   &  &  &          \ding{108}   &  &     \ding{108}   &        \ding{108}   &      \ding{108}   &  &        \ding{108}   &           \ding{108}   &  &  &        \ding{108}   & 1 & 2 & 2 & 1 & 1 &          1 \\
      Math-6   &  &  &  &  &  &  &  &  &       \ding{109}   &  &  &  &  &        \ding{109}   & 1 & 2 & 205 & 1 & 2 &        177 \\
      Math-8   &          \ding{109}   &       \ding{109}   &  &  &  &  &  &  &       \ding{109}   &  &  &  &  &        \ding{109}   & 1 & 1 & 3 & 1 & 3 &          4 \\
     Math-10   &  &  &  &  &  &  &  &  &  &  &  &          \ding{108}   &  &  & 1 & 1 & 1 & 5 & 5 &         17 \\
     Math-11   &  &  &  &  &  &  &  &  &  &  &  &  &          \ding{109}   &  & 1 & 9 & 9 & 27 & 27 &         29 \\
     Math-16   &  &  &  &  &  &  &  &  &  &  &  &  &          \ding{109}   &  & 1 & 1 & 5 & 1 & 1 &          5 \\
     Math-20   &  &  &  &  &  &  &        \ding{109}   &  &       \ding{109}   &  &  &  &  &        \ding{109}   & 1 &         {\bf \color{red}0}  &         {\bf \color{red}0}  & 1 &         {\bf \color{red}0}  &         {\bf \color{red}0}\\
     Math-22   &  &  &  &          \ding{108}   &  &  &  &  &  &  &  &          \ding{108}   &  &  & 1 & 1 & 1 & 1 & 1 &          1 \\
     Math-25   &  &  &  &  &  &     \ding{108}   &  &  &  &  &  &  &  &  &         {\bf \color{red}0}  &         {\bf \color{red}0}  &         {\bf \color{red}0}  &         {\bf \color{red}0}  &         {\bf \color{red}0}  &         {\bf \color{red}0}\\
     Math-28   &          \ding{109}   &       \ding{109}   &            \ding{109}   &  &  &     \ding{109}   &  &  &       \ding{109}   &  &  &  &          \ding{109}   &        \ding{109}   & 6 & 6 & 6 & 13 & 13 &         13 \\
     Math-30   &  &  &  &  &  &  &        \ding{108}   &  &       \ding{108}   &        \ding{108}   &  &          \ding{108}   &  &  & 1 & 4 & 9 & 142 & 161 &        161 \\
     Math-32   &  &       \ding{109}   &  &  &       \ding{109}   &  &        \ding{109}   &    (\ding{108})   &  &  &  &  &  &  & 1 & 3 & 3 & 5 & 5 &          6 \\
     Math-33   &  &  &  &  &       \ding{109}   &  &        \ding{108}   &  &       \ding{108}   &        \ding{108}   &           \ding{108}   &          \ding{108}   &  &        \ding{108}   & 2 & 4 & 31 & 2 & 6 &         44 \\
     Math-34   &  &  &  &          \ding{109}   &  &  &        \ding{108}   &  &  &  &  &          \ding{108}   &  &  & 1 & 1 & 1 & 1 & 3 &          3 \\
     Math-35   &  &  &  &  &  &     \ding{108}   &  &  &  &  &  &  &  &        \ding{108}   &         {\bf \color{red}0}  &         {\bf \color{red}0}  &         {\bf \color{red}0}  & 1 & 3 &          5 \\
     Math-40   &          \ding{109}   &       \ding{109}   &            \ding{109}   &  &       \ding{109}   &  &  &  &  &  &  &  &  &  & 1 & 23 & 24 & 24 & 33 &         34 \\
     Math-41   &  &  &  &  &  &  &  &  &       \ding{108}   &  &  &  &  &        \ding{108}   & 1 & 2 & 6 & 1 & 37 &         49 \\
     Math-42   &  &  &  &  &       \ding{109}   &  &  &  &  &  &  &  &  &  & 1 & 23 & 26 & 3 & 57 &         66 \\
     Math-49   &          \ding{109}   &       \ding{109}   &  &  &       \ding{109}   &  &  &  &  &  &  &  &  &  & 3 & 4 & 7 & 5 & 5 &          7 \\
     Math-50   &          \ding{108}   &       \ding{108}   &            \ding{109}   &          \ding{108}   &       \ding{108}   &  &        \ding{108}   &    (\ding{108})   &       \ding{108}   &  &           \ding{108}   &  &  &        \ding{108}   & 1 & 1 & 1 & 1 & 1 &          1 \\
     Math-53   &          \ding{108}   &  &  &          \ding{108}   &  &  &  &    (\ding{108})   &       \ding{108}   &        \ding{108}   &  &  &  &        \ding{108}   & 1 & 1 & 1 & 1 & 1 &          2 \\
     Math-57   &  &  &            \ding{109}   &  &       \ding{109}   &  &        \ding{108}   &  &       \ding{108}   &        \ding{108}   &  &          \ding{108}   &  &        \ding{108}   & 1 & 1 & 14 & 1 & 4 &          4 \\
     Math-58   &  &  &            \ding{109}   &  &       \ding{109}   &  &        \ding{108}   &  &       \ding{109}   &        \ding{108}   &  &          \ding{108}   &  &  & 6 & 6 & 6 & 223 & 223 &        223 \\
     Math-59   &  &  &  &  &  &  &        \ding{108}   &  &       \ding{108}   &        \ding{108}   &           \ding{108}   &  &  &        \ding{108}   & 1 & 1 & 1 & 1 & 2 &          2 \\
     Math-60   &  &  &  &  &  &  &  &  &       \ding{109}   &  &  &  &  &  & 21 & 21 & 21 & 283 & 283 &        284 \\
     Math-61   &  &  &  &  &  &     \ding{108}   &  &  &  &  &  &  &  &  &         {\bf \color{red}0}  &         {\bf \color{red}0}  &         {\bf \color{red}0}  & 1 & 1 &          1 \\
     Math-63   &  &  &  &  &  &  &        \ding{109}   &  &       \ding{109}   &        \ding{108}   &  &  &          \ding{108}   &        \ding{108}   & 1 & 8 & 8 & 1 & 1 &          1 \\
     Math-65   &  &  &  &  &  &  &  &  &       \ding{109}   &        \ding{108}   &  &  &  &  & 1 & 8 & 9 & 12 & 12 &         15 \\
     Math-69   &  &  &  &  &       \ding{109}   &  &  &  &  &  &  &  &  &  & 1 & 1 & 2 & 1 & 48 &         57 \\
     Math-70   &          \ding{108}   &  &  &          \ding{109}   &  &  &        \ding{108}   &  &       \ding{108}   &        \ding{108}   &           \ding{108}   &          \ding{108}   &          \ding{108}   &        \ding{108}   & 1 & 1 & 1 & 1 & 1 &          1 \\
     Math-71   &          \ding{109}   &  &  &  &       \ding{109}   &  &  &  &       \ding{109}   &  &  &  &  &        \ding{108}   & 1 & 1 & 1 & 1 & 1 &          1 \\
     Math-72   &  &  &  &  &  &  &  &  &  &  &  &  &  &        \ding{109}   & 1 & 3 & 4 & 1 & 1 &          1 \\
     Math-73   &          \ding{108}   &  &  &  &       \ding{109}   &     \ding{109}   &        \ding{109}   &  &  &  &           \ding{109}   &  &  &        \ding{109}   & 1 & 1 & 1 & 1 & 1 &          1 \\
     Math-75   &  &  &  &  &  &  &        \ding{108}   &  &  &        \ding{108}   &  &          \ding{108}   &          \ding{108}   &        \ding{108}   & 1 & 2 & 2 & 1 & 1 &          1 \\
     Math-78   &          \ding{109}   &       \ding{109}   &  &  &       \ding{109}   &  &  &  &       \ding{109}   &  &  &  &  &  & 17 & 21 & 32 & 67 & 67 &        109 \\
     Math-79   &  &  &  &  &  &  &  &  &       \ding{108}   &        \ding{109}   &  &          \ding{108}   &          \ding{108}   &        \ding{108}   & 23 & 23 & 25 & 29 & 29 &         29 \\
     Math-80   &          \ding{109}   &       \ding{109}   &  &  &       \ding{109}   &  &        \ding{109}   &    (\ding{108})   &       \ding{108}   &        \ding{109}   &  &  &          \ding{109}   &        \ding{109}   & 1 & 11 & 18 & 1 & 14 &         14 \\
     Math-81   &          \ding{109}   &       \ding{109}   &            \ding{109}   &  &       \ding{109}   &     \ding{109}   &  &  &       \ding{109}   &        \ding{109}   &  &          \ding{109}   &  &        \ding{109}   & 1 & 1 & 6 & 1 & 1 &         10 \\
     Math-82   &          \ding{109}   &       \ding{109}   &            \ding{108}   &          \ding{109}   &       \ding{109}   &     \ding{108}   &        \ding{108}   &    (\ding{108})   &  &        \ding{109}   &           \ding{108}   &          \ding{108}   &  &        \ding{109}   & 2 & 53 & 60 & 1 & 76 &         84 \\
     Math-84   &          \ding{109}   &       \ding{109}   &            \ding{109}   &  &  &  &  &  &       \ding{109}   &  &  &          \ding{109}   &  &  & 1 & 13 & 30 & 5 & 18 &        134 \\
     Math-85   &          \ding{109}   &       \ding{109}   &            \ding{108}   &  &       \ding{109}   &     \ding{108}   &        \ding{108}   &    (\ding{108})   &       \ding{109}   &        \ding{108}   &           \ding{108}   &          \ding{108}   &  &        \ding{109}   & 1 & 1 & 36 & 11 & 11 &         90 \\
     Math-87   &  &  &  &  &       \ding{109}   &  &  &  &  &  &  &  &  &  & 1 & 99 & 100 & 2 & 109 &        111 \\
     Math-88   &  &  &            \ding{109}   &  &       \ding{109}   &  &  &  &  &  &  &  &  &        \ding{109}   & 1 & 1 & 1 & 1 & 1 &          1 \\
     Math-89   &  &  &  &  &  &     \ding{108}   &  &  &  &  &  &  &          \ding{108}   &  & 1 & 1 & 1 & 1 & 1 &          1 \\
     Math-90   &  &  &  &  &  &     \ding{108}   &  &  &  &  &  &  &  &  & 1 & 1 & 4 & 1 & 1 &          3 \\
     Math-91   &  &  &  &  &  &  &  &  &  &  &  &  &          \ding{108}   &  & 1 & 1 & 2 & 1 & 1 &          1 \\
     Math-93   &  &  &  &  &  &     \ding{108}   &  &  &  &  &  &  &          \ding{109}   &  & 1 & 1 & 2 & 1 & 1 &          2 \\
     Math-94   &  &  &  &  &  &  &  &  &  &  &  &  &          \ding{108}   &  & 1 & 1 & 21 & 1 & 1 &         21 \\
     Math-95   &          \ding{109}   &       \ding{109}   &  &  &  &  &  &  &       \ding{109}   &  &  &  &          \ding{109}   &  & 2 & 2 & 3 & 8 & 11 &         12 \\
     Math-97   &  &  &  &  &       \ding{109}   &     \ding{109}   &  &  &  &  &  &  &  &  & 1 & 1 & 1 & 1 & 1 &          1 \\
     Math-98   &  &  &  &  &  &  &  &  &  &  &  &  &  &        \ding{108}   & 1 & 1 & 6 & 1 & 1 &          4 \\
     Math-99   &  &  &  &  &  &     \ding{108}   &  &  &  &  &  &  &          \ding{109}   &  & 1 & 1 & 1 & 1 & 1 &          4 \\
     Math-104   &  &  &  &  &       \ding{109}   &  &        \ding{109}   &  &  &  &  &  &  &  & 1 &         {\bf \color{red}0}  &         {\bf \color{red}0}  & 1 &         {\bf \color{red}0}  &         {\bf \color{red}0}\\
     Math-105   &  &  &  &  &       \ding{109}   &  &  &      \ding{109}   &  &  &  &  &  &  & 1 & 1 & 1 & 1 & 25 &         25 \\
     \hline  
     Mockito-13   &  &  &  &  &  &  &  &  &  &  &  &  &          \ding{108}   &  & 30 & 30 & 70 & 74 & 74 &        135 \\
     \hline    
     Time-4   &          \ding{109}   &       \ding{109}   &  &  &  &  &        \ding{108}   &  &       \ding{109}   &  &           \ding{109}   &  &  &  & 36 & 36 & 208 & 6 & 6 &         31 \\
     Time-7   &  &  &  &  &  &  &  &  &  &  &  &  &  &        \ding{108}   & 48 & 48 & 51 & 10 & 10 &         14 \\
     Time-11   &          \ding{109}   &       \ding{109}   &            \ding{109}   &  &       \ding{109}   &  &        \ding{109}   &  &       \ding{109}   &  &  &  &  &  & 4 &         {\bf \color{red}0}  &         {\bf \color{red}0}  & 51 &         {\bf \color{red}0}  &         {\bf \color{red}0}\\
     Time-14   &  &  &  &  &  &  &  &  &       \ding{109}   &  &  &  &  &  & 4 & 4 & 7 & 2 & 2 &          3 \\
     Time-15   &  &  &  &  &  &     \ding{108}   &        \ding{108}   &  &  &  &  &  &  &  & 1 & 1 & 115 & 1 & 1 &          2 \\
     Time-17   &  &  &  &  &  &  &  &  &       \ding{109}   &  &  &  &  &  & 5 & 5 & 5 & 5 & 5 &          5 \\
     Time-19   &  &  &  &          \ding{109}   &  &  &  &  &  &  &  &          \ding{108}   &  &  & 5 & 449 & 449 & 104 & 620 &        620 \\
\hline
\end{tabular}

			}
		\end{minipage}
	\vspace{2mm}
	\end{tabular}
	{\footnotesize  $\ast$ \ding{108} indicates that the bug is correctly fixed and \ding{109} indicates that the generated patch is plausible but not correct. (\ding{108}) indicates that a correct patch is generated, but is not the first plausible patch to be generated''. ``{\bf \textcolor{red}{0}}'' means that the bug cannot be localized by the corresponding FL tool with the corresponding ranking metric in the corresponding granularity.}
\end{table*}

We propose to compute the distributions of positions across subsets of bugs for checking correlations between the localization ranking positions
 and the ability of APR systems to fix the bugs. Thus, we normalize bug localization positions by computing reciprocal positions based on the following formula:
\vspace{-1.5mm}
\begin{equation}
\vspace{-1.5mm}
\scriptsize
	\begin{array}{l}
    	{Reciprocal_{pos}}(bug_{pos}) =
    	\begin{cases}
			0, &  \text{if } bug_{pos} = 0; \\
 			\frac{1.0}{bug_{pos}}, & otherwise.
 		\end{cases}
 	\end{array}
\label{eq:normalization}
\end{equation}
where $bug_{pos}$ refers to the position of the actual bug location\footnote{If several lines are concerned by the bugs,
we consider the first time any of these lines appear as the bug position (cf. Section~\ref{sec:bg}).} in the ranked list of suspicious locations reported by the FL step.
If the bug location can be found in the higher position of the ranked list,
the value of $Reciprocal_{pos}$ is closer to 1.
Similarly, the value of $Reciprocal_{pos}$ trends to 0 when the bug location is at lower positions in the list of suspicious locations.
This value is set to 0 when the bug cannot be localized by the FL tool (i.e., $bug_{pos} = 0$).
In addition, for the purpose of our experiments, we consider three sub-classes of bugs:
\begin{itemize}[leftmargin=*]
	\item {\em correctly fixed bugs}:  these are bugs for which a correct patch has been provided by at least one APR tool.
	\item {\em overfitting-fixed bugs}: these are bugs for which one or more plausible patch has been generated, although none has been found to be correct.
	\item {\em unfixed bugs}: these are bugs for which no plausible patch has ever been generated by any APR system. Due to space limitation, localization data for these bugs are only available in the replication package.
\end{itemize}

\begin{figure}[!t]
    \centering
    \includegraphics[width=0.8\linewidth]{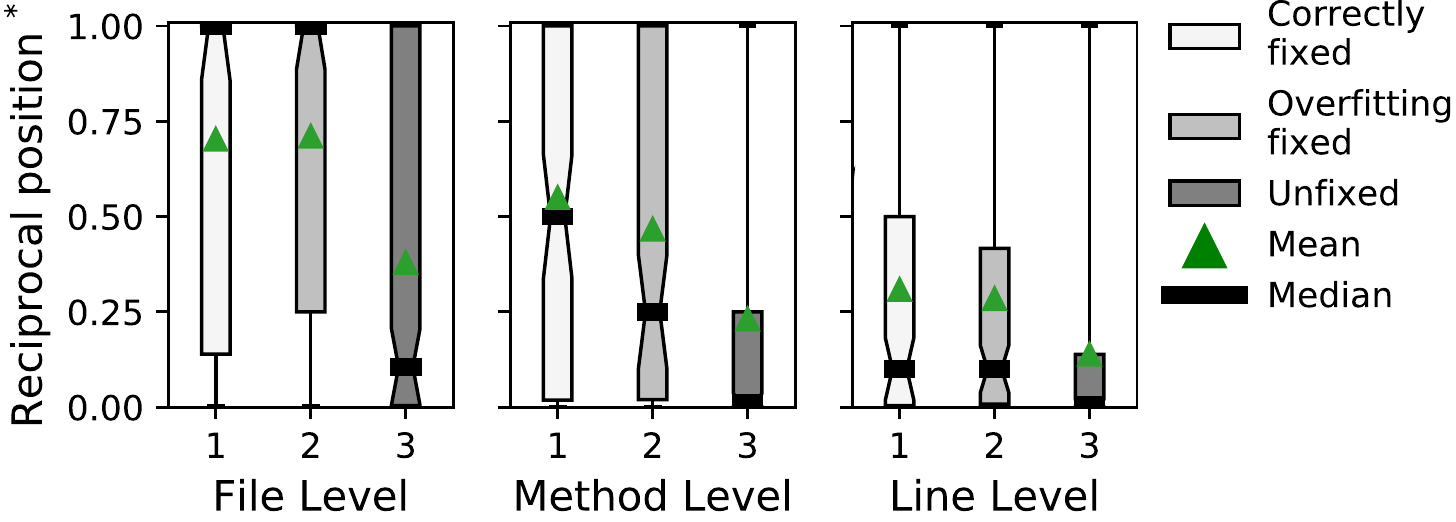}
    \caption{Distribution of reciprocal positions of actual bug locations among the ranked list of suspicious locations.}
        \label{fig:distOfNormalization}
\end{figure}

Figure~\ref{fig:distOfNormalization}\footnote{The bug positions before being reciprocated shown in the figure are localized by GZoltar 0.1.1 with Ochiai.} shows the distribution of reciprocal positions for the three classes of bugs at the file, method, line granularity of fault locality.
It clearly appears that correctly-fixed bugs are more accurately localized than others: i.e., their location precisions are higher in the ranked list of suspicious locations by FL tools. On the other hand, unfixed bugs tend to be those that are poorly localized: even at the file level, FL tool show low performance in localizing such bugs.

\vspace{2mm}
\find{{\bf RQ2}$\blacktriangleright$APR tools are prone to correctly fix the subset of Defects4J bugs that can be accurately localized.}

We further observe from the data in Table~\ref{tab:results.apr} that a few
APR systems report patches for some bugs even though they cannot be localized
(at the line level) with the configuration of Ochiai/GZoltar 0.1.1.
There are various justifications to this phenomenon:
\begin{itemize}[leftmargin=*]
	\item {\bf Improved version of the fault localization step} -
	{\em Chart-20} cannot be localized with GZoltar 0.1.1 and Ochiai,
	but is reported to be fixed by tools such as SimFix and ssFix.
	Our investigations show that SimFix has used a recent version of GZoltar (1.6.0), which is capable of localizing {\em Chart-20} among other bugs that were not localizable.
	ssFix on the other hand indeed uses GZoltar 0.1.1 but do not consider only the results of the FL tool:
	statements in the stack trace of crashed programs are also considered as potential fault locations.
	\item {\bf Targeted localization} - HDRepair can fix {\em Lang-6},
	which is not localized with Ochiai/GZoltar 0.1.1,
	because this APR system assumes that the faulty method is known,
	and thus directly ranks the restricted set of statements in this method.
	\item {\bf Coarse-grained repair} - LSRepair can fix four bugs which cannot be localized
	at the line granularity. This is due to the fact that LSRepair requires
	only fault localization at the method level, which is not a bias per se.
	\item {\bf Non-explicit fault localization process} - SketchFix, JAID, and ELIXIR correctly
	fix some bugs that are not localized under the proposed configuration. Unfortunately,
	besides the lack of details in their associated research reports,
	the source code of these tools was not made available for further investigation.
	{\em Chart-8} is another example that is not localizable by using Ochiai/GZoltar 0.1.1.
	This specific un-localizability problem was recently raised by
	Yuan and Banzhaf~\cite{yuan2017arja} as well as Martinez et al.~\cite{martinez2017automatic}.
	Nevertheless, CapGen, ELIXIR and SketchFix are reported to have fixed this bug.
\end{itemize}

\find{{\bf RQ3}$\blacktriangleright$APR systems do not fully disclose their fault localization tuning parameters, thus preventing reliable replication and comparisons.}

Given the bias that can be introduced by unlocalizable bugs being fixed by specific tweaking,
which are not clearly outlined by the authors,
we propose to count the numbers of bugs that are fixed by
APR systems among those bugs that are known to be localizable.
Table~\ref{tab:localizableComparison} thus represents an updated version of
Table~\ref{tab:comparison} where performance can be compared on the same basis.
To illustrate the differences between the two comparison tables, we compute three scores: (1) {\bf NPFB}: number of plausibly-fixed bugs, (2) {\bf NCFB}: number of correctly-fixed bugs, and (3) {\bf P$^3$C}: probability of plausible patch correctness.

\begin{figure}[!h]
	\hfill
	\subfigure[\# of plausibly-fixed bugs.]{\includegraphics[width=0.47\linewidth]{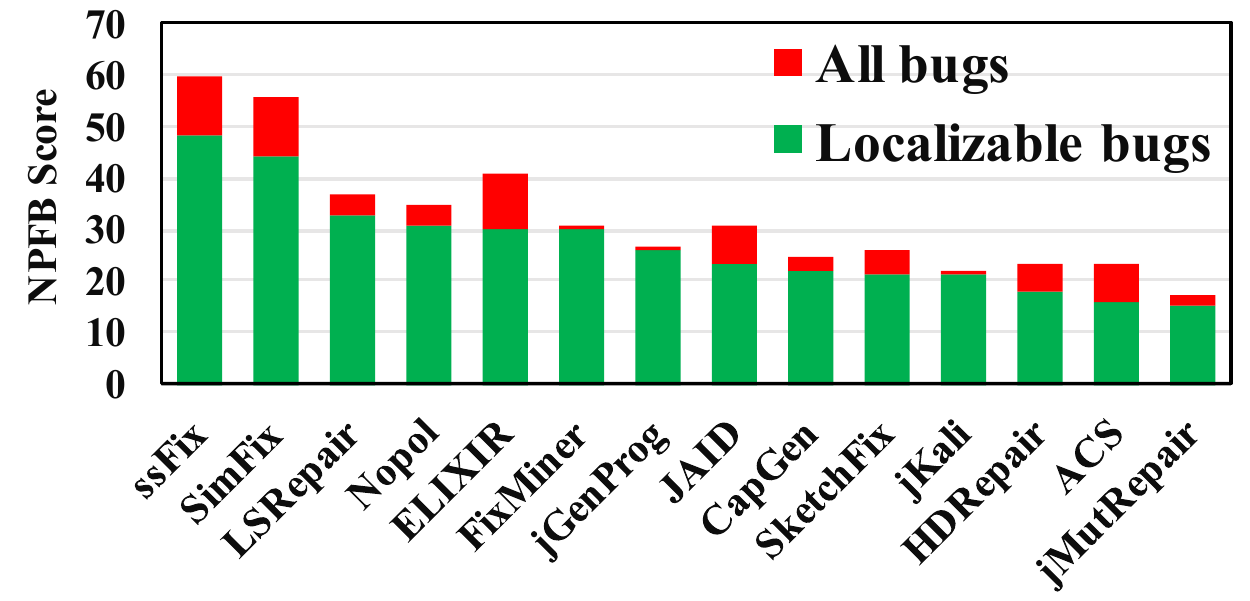}
	\label{fig:npfb}}
	\hfill
	\subfigure[\# of correctly-fixed bugs.]{\includegraphics[width=0.47\linewidth]{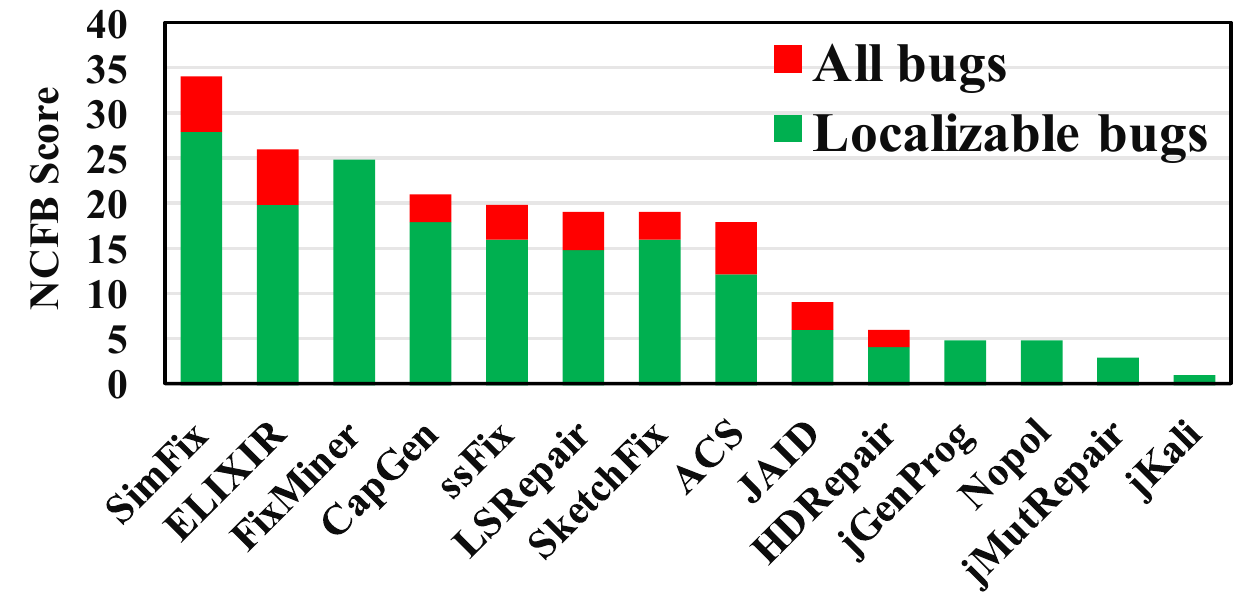}
	\label{fig:ncfb}}
	\hfill
	\vspace{-1mm}
	\caption{Number of fixed bugs among all bugs vs. localizable bugs.}
	\label{fig:multiScores}
\end{figure}

Figures~\ref{fig:npfb} and \ref{fig:ncfb} illustrate the differences in respectively NPFB and NCFB scores when considering all bugs vs only localizable bugs. We note that all tools may produce some plausible patches that are plausible even for non-localizable bugs. This finding suggests that the test cases in Defects4J are insufficient since it is possible for APR systems to change non-faulty code locations and still produce patches that make the faulty program pass all test cases.
On the other hand, five APR systems cannot produce any correct patches for bugs that are not localizable. ACS, ELIXIX and SimFix can correctly fix bugs that are not localized with GZoltar 0.1.1, suggesting extra impact with an improved version of the fault localization step. On the other hand, LSRepair can fix bugs that are not localized at the line level because method level fault localization is sufficient for its execution.

Finally, Table~\ref{tab:p3c} establishes the re-ranking of APR systems in terms of the {\bf P$^3$C} scores when focusing on localizable bugs. When focusing on localizable bugs, state-of-the-art APR systems can correctly overall fix fewer bugs than reported in the literature.

\begin{table}[!h]
	\centering
   	\scriptsize
   	\caption{Adjusted Probability of Plausible Patch Correctness.}
    \label{tab:p3c}
   		\begin{tabular}{|c|r|c|c|r|}
			\hline
			\multicolumn{2}{|c|}{\bf All} & &\multicolumn{2}{c|}{\bf Localizable} \\
			\hline
			P$^3$C & Rank & Tool & P$^3$C & Rank \\
	 		\hline
		 	84.0	& 1	&	CapGen		& 81.8 & $\downarrow$ 2\\
	 		80.6	& 2	&	FixMiner	& 83.3 & $\uparrow$ 1\\
	 		78.3	& 3	&	ACS 		& 75.0 & $\downarrow$ 4\\
	 		73.1	& 4	&	SketchFix	& 76.2 & $\uparrow$	3\\
	 		63.4	& 5	&	ELIXIR		& 66.7 &  5\\
	 		60.7	& 6	&	SimFix		& 63.6 & 6\\
	 		51.4	& 7	&	LSRepair	& 45.5 & 7\\
	 		33.3	& 8	&	ssFix		& 33.3 & 8\\
	 		29.0	& 9	&	JAID		& 26.1 & 	9\\
	 		26.1	& 10&	HDRepair	& 22.2 & 	10\\
	 		18.5	& 11&	jGenProg	& 19.2 & $\downarrow$ 12\\
	 		17.6	& 12&	jMutRepair	& 20.0 & $\uparrow$ 11\\
	 		14.3	& 13&	Nopol		& 16.1 & 			13\\
	 		 4.5 	& 14&	jKali		& 4.8 & 			14\\
	 		\hline
		\end{tabular}
\end{table}

\begin{table*}[!t]
	\scriptsize
	\centering
	\setlength\tabcolsep{2pt}
	\caption{Number of localizable bugs (with GZoltar 0.1.1 and Ochiai) fixed by different APR tools.}
	\label{tab:localizableComparison}
    	\begin{threeparttable}
			\begin{tabular}{l|c|C{10mm}|c|c|C{10mm}|C{10mm}|c|C{10mm}|C{10mm}|c|c|c|c|C{10mm}}
			\toprule
			{\bf Proj.} & {\bf jGenProg} & {\bf jKali} & {\bf jMutRepair} & {\bf HDRepair} & {\bf Nopol} & {\bf ACS} & {\bf ELIXIR} & {\bf JAID} & {\bf ssFix} & {\bf CapGen} & {\bf SketchFix} & {\bf FixMiner} & {\bf LSRepair} & {\bf SimFix} \\
	        \hline
 Chart   & 0/7 & 0/6 & 1/4 & 0/1 & 1/6 & 2/2 & 3/6 & 2/4 & 2/6 & 3/3 & 4/6 & 5/7 & 3/8 & 3/6 \\
 Closure & 0/0 & 0/0 & 0/0 & 0/6 & 0/0 & 0/0 & 0/0 & 3/8 & 2/8 & 0/0 & 3/4 & 5/5 & 0/0 & 6/6 \\
 Lang    & 0/0 & 0/0 & 0/0 & 0/3 & 3/5 & 0/0 & 3/5 & 0/3 & 2/6 & 3/3 & 2/2 & 2/3 & 4/10 & 5/8 \\
 Math    & 5/18 & 1/14 & 2/11 & 4/7 & 1/20 & 9/13 & 12/17 & 1/8 & 10/25 & 12/16 & 7/8 & 12/14 & 7/14 & 13/23 \\
 Mockito & 0/0 & 0/0 & 0/0 & 0/0 & 0/0 & 0/0 & 0/0 & 0/0 & 0/0 & 0/0 & 0/0 & 0/0 & 1/1 & 0/0 \\
 Time    & 0/1 & 0/1 & 0/0 & 0/1 & 0/0 & 1/1 & 2/2 & 0/0 & 0/3 & 0/0 & 0/1 & 1/1 & 0/0 & 1/1 \\
			\hline
		    Total &	5/26 & 1/21 & 3/15 & 4/18 & 5/31 & 12/16 & 20/30 & 6/23 & 16/48 & 18/22 & 16/21 & 25/30 & 15/33 & 28/44 \\
			\hline
			\hline
\cellcolor{grey}{Total$^\ast$ (all bugs)} & \cellcolor{grey}{5/27} & \cellcolor{grey}{1/22} & \cellcolor{grey}{3/17} & \cellcolor{grey}{6/23} & \cellcolor{grey}{5/35} & \cellcolor{grey}{18/23} & \cellcolor{grey}{26/41} & \cellcolor{grey}{9/31} & \cellcolor{grey}{20/60} & \cellcolor{grey}{21/25} & \cellcolor{grey}{19/26} & \cellcolor{grey}{25/31} & \cellcolor{grey}{19/37} & \cellcolor{grey}{34/56} \\
			\hline
			\hline
P(\%) 	&	 19.2 &	 4.8 & 20.0 & 22.2 & 16.1 & 75.0 & 66.7 & 26.1 & 33.3 & 81.8 & 76.2 & 83.3 & 45.5 & 63.6 \\
			\hline
			\hline
\cellcolor{grey}{P(\%)$^\ast$ (all bugs)} & \cellcolor{grey}{18.52} & \cellcolor{grey}{4.55} & \cellcolor{grey}{17.65} & \cellcolor{grey}{26.09} & \cellcolor{grey}{14.29} & \cellcolor{grey}{78.26} & \cellcolor{grey}{63.41} & \cellcolor{grey}{29.03} & \cellcolor{grey}{33.33} & \cellcolor{grey}{84.00} & \cellcolor{grey}{73.08} & \cellcolor{grey}{80.65} & \cellcolor{grey}{51.35} & \cellcolor{grey}{60.71}  \\
			\bottomrule
		\end{tabular}
		{\footnotesize {\bf $^\ast$}Greyed-out rows are copied from Table~\ref{tab:comparison} (i.e., numbers reported in the literature) to ease comparison with the numbers of localizable bugs that are fixed.}
		\end{threeparttable}

\end{table*}

\subsection{Evaluating \toolname with Specific FL Configurations}
\toolname is an open-source APR system that we have built to provide a baseline for comparisons of different FL configurations.
We evaluate its performance against the Defects4J benchmark with
the following four different configurations of the fault localization step:
\begin{enumerate}
	\item {\bf Normal\_FL} gives a ranked list of suspicious code locations identical as reported by a given FL tool.
	\item {\bf File\_Assumption} assumes that the faulty code files are known. Suspicious code locations from Normal\_FL are then filtered accordingly. In other words, locations in the known buggy files are selected and locations in other files are ignored.
	\item {\bf Method\_Assumption} assumes that the faulty methods are known (the same assumption with~\cite{xuan2016history}). Only locations in the known methods are selected and locations in other methods are ignored.
	\item {\bf Line\_Assumption} assumes that the faulty code lines are known. No fault localization is then used.
\end{enumerate}
These configurations have an order with respect to a potential size of the search space.
Conceptually, the relationships between them hold {\footnotesize $P( |Normal\_FL|) \leq P(|File\_Assumption|)  \leq  P(|Method\_Assumption|)  \leq  P(|Line\_Assumption|)$},
if we consider each configuration as producing a set of suspicious locations,
where {\footnotesize $P(|\ast|)$} is the probability that the relevant fault locations are included in the suspicious list.

To facilitate comparison with existing repair systems, we leverage the standard GZoltar 0.1.1 and Ochiai in the following experiments. For each bug, we apply \toolname at most three hours (wall-clock time);
we assume that it fails to fix a given bug if it takes more than three hours. We set this value according to the experimental setup of Astor~\cite{martinez2016astor}.
Table~\ref{tab:fixedBugs} summarizes the number of bugs fixed by \toolname with the different FL configurations.

As shown in Table~\ref{tab:fixedBugs}, \toolname can fix its maximum number of bugs when the accurate fault locations are provided (i.e., with {\em Line\_Assumption}).
With this assumption, \toolname can correctly fix 36 bugs in Defects4J, a record performance in the literature (not accounting for the bias in the fault localization step).

\begin{table}[!h]
    \centering
    \scriptsize
    \caption{\# of Bugs fixed by \toolname.}
    \label{tab:fixedBugs}
    \resizebox{1.00\linewidth}{!}
    {
    \begin{threeparttable}
    \begin{tabular}{l|c|c|c|c|c|c|c}
        \toprule
        {\bf FL Configuration}   & {\bf Chart (C)} & {\bf Closure (Cl)} & {\bf Lang (L)} & {\bf Math (M)} & {\bf Mockito (Moc)} & {\bf Time (T)} & {\bf Total} \\
        \hline
        Normal\_FL          & 3/10  &   5/9   & 1/8  & 7/18 & 1/2     & 1/2  & 18/49\\
        \hline
        File\_Assumption       & 4/7   &   6/13  & 1/8  & 7/15 & 2/2     & 2/3  & 22/48\\
        \hline
         Method\_Assumption      & 4/6   &   7/16  & 1/7  & 7/15 & 2/2     & 2/3  & 23/49\\
        \hline
        Line\_Assumption  & 7/8   &  11/16  & 4/9  & 9/16 & 2/2     & 3/4  & 36/55\\
        \bottomrule
    \end{tabular}
    \end{threeparttable}
    }
\end{table}

\begin{figure}[!h]
	\centering
    \includegraphics[width=0.9\linewidth]{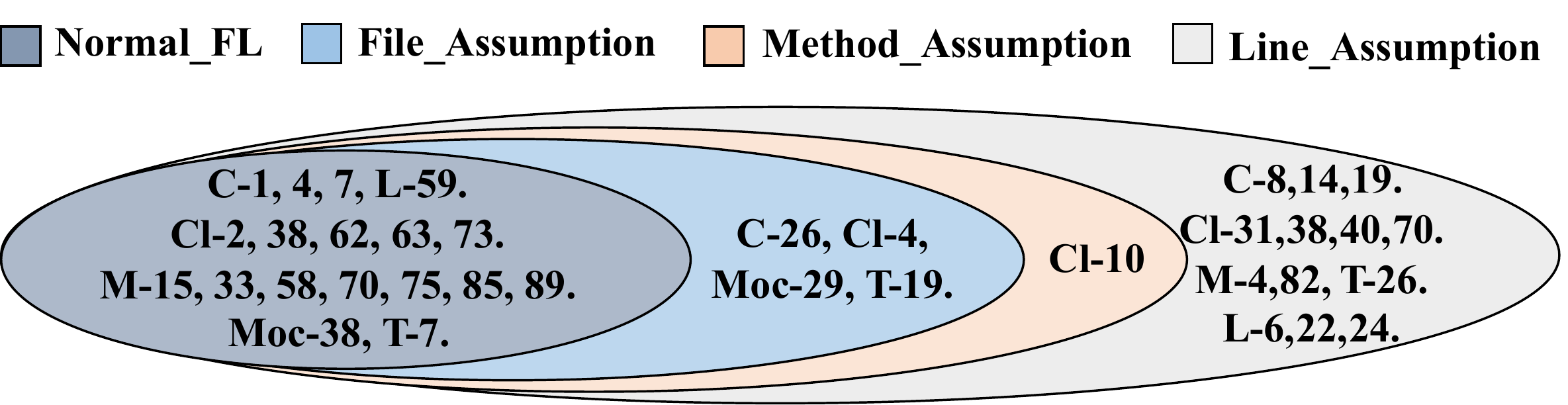}
    \caption{Bugs correctly fixed by \toolname with four configurations.}
    \label{fig:fixedBugs}
\end{figure}

Figure~\ref{fig:fixedBugs} further details which bugs are fixed in the different configurations.
First, we note that all bugs fixed with a given localization configuration are also fixed by any of the relatively more accurate fault localization configurations.
Thus, with the {\em File\_Assumption} configuration, \toolname can fix not only all bugs that were already fixed with the {\em Normal\_FL} configuration but also can now fix four more bugs.
By examining the case of those four bugs, we figure out that, in the case of two bugs
(i.e., {\em Cl-4} and {\em T-19}), the faulty locations were ranked very low in {\em Normal\_FL},
leading to an execution stop due to timeout.
For the remaining two bugs (i.e., {\em C-26} and {\em Moc-29}), however,
in {\em Normal\_FL}, \toolname is led to consider first some irrelevant suspicious statements that made
\toolname to generate plausible patches that are not correct.
Given that the repair process stops when a plausible patch is produced,
there is no opportunity with {\em Normal\_FL} to try all suspicious statements.

 When filtering the set of suspicious locations with {\em Method\_Assumption}, \toolname can fix one more bug (i.e., {\em Cl-10}),
 which could not be fixed by other two less confined FL configurations (i.e., {\em Normal\_FL} and {\em File\_Assumption}) before the time-out.
 Finally, when assuming that the fault locations are known (i.e., {\em Line\_Assumption}), \toolname can further fix 13 bugs.
 These could not be fixed in other three less confined configurations.
 Among the 13 bugs, seven bugs (i.e., {\em C-8}, {\em Cl-40}, {\em Cl-70}, {\em L-6}, {\em L-22}, {\em L-24}, and {\em T-26}) are not even localizable using Ochiai/GZoltar 0.1.1; two bugs (i.e., {\em Cl-18} and {\em Cl-70}) are not fixed due to execution timeout; one bug (i.e., {\em M-82}) is not fixed in other three configurations since the proposed plausible patches are incorrect; three bugs (i.e., {\em C-14}, {\em C-19} and {\em M-4}) are partially fixed in the other three FL configurations since they have several faulty code fragments.

 \find{{\bf RQ3}$\blacktriangleright$Accuracy of fault localization has a direct and substantial impact on the performance of APR repair pipelines.}

We examine 
the bug {\em Chart-14} from the Defects4J dataset, which involves four fault code locations~\cite{chart14}.
If we regard those as four sub-bugs, each one can be correctly detected and fixed by \toolname using the {\em Normal\_FL} configuration.
However, if the exact faulty statements are unknown, \toolname (as current APR tools) iteratively mutates suspicious statements one by one in the ranked list. Even if any one of them
is correctly fixed, there are still three failed tests, meaning that the generated patch (even if was a correct patch) will not even be considered as a plausible patch.

Considering a patch that partially passes some previously-failing test cases (without introducing new failing test cases) may nevertheless be harmful as it can prevent the generation of a fully correct patch. For example, {\em Chart-4} is a single-location bug that makes 22 test cases fail~\cite{chart4}.
Before generating the correct patch, \toolname had generated patches that made the program pass subsets of the test cases.

Other bugs, such as {\em Math-72}, on the other hand include multiple faulty locations that fail on the same test case.
Although \toolname could generate correct patches for each faulty location, the fix process of \toolname prevents a full fix of this bug.
If the test suite can be automatically augmented with differentiating test cases for each fault location, an APR system would be more successful as suggested in~\cite{yang2017better}.

\vspace{1mm}
\find{{\bf RQ3}$\blacktriangleright$APR researchers must investigate the trade-off between fixing multi-locations bugs versus bugs failing multiple test cases.}

\vspace{-1.5mm}
\section{Discussion}
\label{sec:dis}

Our study draws a number of conclusions that we reformulate into guidelines for assessing APR systems. We further enumerate the associated threats to validity before discussing the related work.

\subsection{APR Assessment Guidelines}
\begin{itemize}[leftmargin=*]
	\item {\bf Full disclosure of FL parameters.} Given that many APR systems do not release their source code, it is important that the experimental reports clearly indicate the protocol used for fault localization. Preferably, authors should strive to assess their performance under a standard and replicable configuration of fault localization.
	\item {\bf Qualification of APR performance.} To ensure that novel approaches to APR are indeed improving over the state-of-the-art, authors must qualify the performance gain brought by the different ingredients of their approaches.
	\item {\bf Patch generation step vs Repair pipeline.} There are two distinct directions of repair benchmarking that APR researchers should consider. In the first, a novel contribution to the patch generation problem must be assessed directly by  assuming a perfect fault localization.  In the second, for ensuring realistic assessment w.r.t. industry adoption, the full pipeline should be tested with no assumptions on fault localization accuracy.
	\item {\bf Sensitivity of search space.} Given that fault localization produces a ranked list of suspicious locations, it is essential to characterize whether exact locations are strictly necessary for the APR approach to generate the correct patches. For example, an APR system may not focus only on a suspected location but on the context around this location. APR approaches may also use heuristics to curate the FL results.
\end{itemize}

\subsection{Threats to Validity}
A threat to external validity of our study is that we focus on the localizability of bugs in the Defects4J dataset, which target Java code and may not include sufficient test cases. This threat is however limited given that we investigate performance differences.
Threats to internal validation include the use of a single automatic testing framework, namely GZoltar (Not all APR systems in the literature use it to localize faults.), and the selection of the 14 state-of-the-start APR systems. These threats are mitigated by the fact that we ensured that these choices are common among the APR literature.

\subsection{Related Work}
The software development practice is increasingly accepting generated patches~\cite{koyuncu2017impact}. Recently, various  directions in the literature have explored to contribute to the advancement of automated program repair~\cite{jones2005empirical, abreu2007accuracy,abreu2009practical,janssen2009zoltar,thung2012faults,wong2016survey,perez2017test,pearson2017evaluating}. We now discuss the few related studies that attempt to investigate fault localization in relationship with APR.

Qi et al.~\cite{qi2013using} have evaluated the effectiveness of FL tools by using APR performance as a proxy.
Their study proposed the NCP score~\cite{qi2013using}
as the  effectiveness metric.
The results show that a specific FL ranking metric (Jaccard~\cite{chen2002pinpoint}) outperforms
other metrics. Our study, however, reveals that the common technique used in APR is still Ochiai.
Yang et al.~\cite{yang2017empirical} studied the usage of FL techniques in APR systems by investigating
two different algorithms of how to interpret the results of FL techniques:
(1) the rank-first algorithm based on suspiciousness rankings of statements, and (2) the suspiciousness-first algorithm based on suspiciousness scores of statements.
They ran Nopol~\cite{xuan2017nopol} to compare NCP scores, repair time, and patch diversity of the two algorithms. The study concludes that the suspiciousness-first algorithm is more effective for APR systems.
The above two studies, however, do not consider whether the patches generated by APR tools are correct or plausible while our study examines how FL techniques affect the quality of patches generated by APR systems.

The literature also includes work on the impact of the fault space, although it does not clarify how FL tools affect the performance of APR systems.
Wen et al.~\cite{wen2017empirical} investigated the influence of the fault space on the success of finding correct patches by the APR tool.
The fault space is defined as a ranked list of suspicious entities in a program.
They examined both plausible and correct patches. However, their work is limited to evaluating a single APR tool, GenProg~\cite{claire2012genprog} and a single FL technique, Ochiai~\cite{abreu2007accuracy} while our study evaluates and compares 14 different APR systems.
Our study further considers the exact location of faults and its correlation with the possibility of generating plausible patches. Finally, our study targets unveiling biases among APR systems.

To the best of our knowledge, our work is the first time to systematically study to what extent FL techniques impact the performance of automated program repair pipeline.

\section{Conclusion}
\label{sec:conc}
The momentum of research in automated program repair is a decisive opportunity for the software engineering research community. 
Every couple of months, a new APR system is proposed in a race to fix more bugs automatically. Unfortunately, validation of these systems
often have only the dataset in common: important parameters such as the fault localization settings are eluded, leading to biased comparisons among the state-of-the-art. Our investigations into these biases call for new guidelines for assessing and reporting on the performance of APR systems. In particular, our replication package includes a full dissection of the Defects4J benchmark in terms of fault localization, a light-weight and tuneable fault localization toolkit, as well as a baseline Java APR system to encourage fair and reproducible experiments.

\vspace{2mm}
\noindent
\section*{Acknowledgements}{
This work is supported by the Fonds National de la Recherche (FNR), Luxembourg, under projects RECOMMEND 15/IS/10449467 and
FIXPATTERN C15/IS/9964569.}

\balance
\bibliographystyle{IEEEtran}
\bibliography{bib/references}

\end{document}